\documentclass[pre,showpacs,twocolumn]{revtex4}
\def\figwidth{0.9\linewidth}
\usepackage{amsmath}
\usepackage{amsfonts}
\usepackage[final]{graphicx}

\let\bs\boldsymbol

\def\sgn{\qopname\relax{no}{sgn}}

\begin{document}
\title{Effect of Composition Changes on the Structural Relaxation
       of a Binary Mixture}
\date{\today}
\author{W.~G\"otze}
\author{Th.~Voigtmann}
\affiliation{Physik-Department, Technische Universit\"at M\"unchen,
             85747 Garching, Germany}

\begin{abstract}
Within the mode-coupling theory for idealized glass transitions, we study
the evolution of structural relaxation in binary mixtures of hard spheres with
size ratios $\delta$ of the two components varying between $0.5$ and
$1.0$. We find two scenarios for the glassy dynamics. For small
size disparity, the mixing yields a slight extension of the glass
regime. For larger size disparity, a plasticization effect
is obtained, leading to a stabilization of the liquid due to mixing. For
all $\delta$, a decrease of the elastic moduli at the transition due to
mixing is predicted. A stiffening of the glass structure is
found as is reflected by the increase of the Debye-Waller factors
at the transition points. The critical
amplitudes for density fluctuations at small and intermediate wave vectors
decrease upon mixing, and thus the universal formulas for the
relaxation near the plateau values describe a
slowing down of the dynamics upon mixing for the first step of the
two-step relaxation scenario.
The results explain the qualitative features of mixing
effects reported by Williams and van~Megen [Phys.~Rev.~E \textbf{64},
041502 (2001)] for dynamical light-scattering measurements on binary
mixtures of hard-sphere-like colloids with size ratio $\delta=0.6$.
\end{abstract}

\maketitle

\section{Introduction}

The study of glass-transition phenomena in dense colloidal suspensions
has received much attention during the past years.  Such systems are well
suited for a test of theories since the particles'
properties can be tuned within a broad range.
In particular, one can produce mixtures with particles of different sizes
and observe effects of changing composition or size ratio. Recent
experimental work of Williams and van~Megen \cite{Williams2001b} has
shown that even in the simplest such systems, namely binary hard-sphere
mixtures (HSM), interesting mixing phenomena appear for the dynamics close to
the glass transition. Three effects
have been reported going over from a one-component to a binary system
containing up to 20\% by volume of smaller spheres:
(i) a shift of the glass transition to higher packing fractions,
(ii) an increase in the plateau values of the correlation
functions at intermediate times, connected to an increase in the glass
form factors, and
(iii) a slowing down of the initial part of the relaxation towards this
plateau.

In this paper, mixing effects in binary HSM are investigated in
the framework of the mode-coupling theory of the idealized glass transition
(MCT). The study of glass-transition phenomena in
colloidal suspensions that are good realizations of one-component
hard-sphere systems has revealed that MCT describes much of the experimental
facts in these cases \cite{Megen1993b,Megen1995}.
MCT makes general predictions for all glass-forming systems, independent
of their underlying microscopic properties, be they one- or multi-component
systems. Thus, a universal glass-transition scenario has been established,
involving
scaling laws and power-law variations of time scales. These properties have
been found in many, not only colloidal, systems, as reviewed in
Refs.~\cite{Goetze1999,Kob1999}.
But MCT is also able to derive detailed results depending on the
specific interactions of a system. The aforementioned
hard-sphere colloids are a paradigmatic example for which, among other things,
the wave-vector dependence of the Debye-Waller factors in the glass state has
been evaluated and compared with results from scattering experiments
\cite{Megen1995}. The quantitative study of model systems allows one to predict
general, while nonuniversal, trends that arise in certain classes of glass
formers. Such project has been
carried out for molecular liquids, where the known differences of
reorientational relaxation for angular momentum $\ell=1$ and $\ell=2$ could be
explained \cite{Goetze2000c,Chong2001}. The
work presented here in a similar way aims to explain the general trends
occurring in a mixture when changing its composition or the size disparity
of its constituents.
Our discussion, motivated by the cited light-scattering experiments
\cite{Williams2001b}, focuses on binary HSM with not too
large size disparity in the species, close to the glass-transition density.

For a derivation of the MCT for mixtures, the reader is referred to
Ref.~\cite{Goetze1987b}. The theory has already been applied to
analyze computer-simulation data for a binary soft-sphere mixture
\cite{Barrat1990b}, a binary Lennard-Jones mixture \cite{Nauroth1997,Kob2002},
a molecular-dynamics model of a silica melt \cite{Sciortino2001}, and of a
two-component metallic melt \cite{Mutiara2001}.
Also, properties of binary HSM in the limit of
large size disparity \cite{Bosse1987,Thakur1991a,Thakur1991b,Bosse1995,
Bosse1997b} and of charged hard spheres, particularly in their low-density
region \cite{Thakur1990,Bosse1998,Chen1999}, have been studied in the
framework of MCT.
Mixing effects in a binary HSM have been
addressed recently using a standard liquid-state mode-coupling approximation,
albeit for states of such low density that glassy
dynamics does not occur \cite{Srinivas2001,Mukherjee2001}.
MCT equations for mixtures have been derived recently within a
nonlinear-hydrodynamics theory \cite{Harbola2002}. The found equations are very
different from the ones analyzed here, and a connection of their implications
with the light-scattering data \cite{Williams2001b} was not discussed.

The paper is organized as follows. In Sec.~\ref{sec.eom}, we summarize the
basic formulas specifying the model under study. Sections~\ref{sec.phase} and
\ref{sec.nep} discuss our results for the fluid-glass transition diagram and
the glass-form factors, respectively. We demonstrate in Sec.~\ref{sec.dyn}
that these lead to two qualitatively different scenarios for the
dynamics close to the glass transition. Section~\ref{sec.conc}
summarizes the results.

\section{Definition of the Model}\label{sec.eom}

\subsection{General Equations of Motion}

A classical $S$-component fluid of $N$ spherical particles shall
be considered.
The fluctuations of the partial number-densities
shall be denoted as $\varrho_\alpha(\vec q)=\sum_k\exp[i\vec q\vec r_k
^{\,(\alpha)}]/\sqrt{N}, \alpha=1,2,\ldots,S$,
where the sum runs over all $N_\alpha$ particle positions $\vec r_k
^{\,(\alpha)}$ belonging to species $\alpha$.
From this, the partial density correlators are constructed,
written $\Phi_{\alpha\beta}(q,t)=\langle\varrho_\alpha(\vec q)|
\varrho_\beta(\vec q,t)\rangle$. Here, $\langle A|B\rangle=\langle\delta A^*
\delta B\rangle$ with $\delta A=A-\langle A\rangle$ denotes a scalar
product in the space of dynamical variables. Angle brackets indicate
canonical averaging for temperature $T$.
The time evolution is generated by a Liouvillian ${\mathcal L}$:
$\varrho_\alpha(\vec q,t)=\exp[i{\mathcal L}t] \varrho_\alpha(\vec q)$.
Since the $\varrho_\alpha$ are spatial
Fourier transforms of a real density variable in an isotropic, translational
invariant system,
$\Phi_{\alpha\beta}(q,t)$ is real, even in $t$, and it depends on the wave
vector only through $q=|\vec q|$.
An evaluation of the density correlators $\Phi_{\alpha\beta}(q,t)$ is the
major aim of this paper.

The starting point of the theory is the exact reformulation of the equations
of motion using the Zwanzig-Mori technique. Considering the limit of
a system of colloidal particles undergoing Brownian dynamics, this equation
reads
\begin{subequations}\label{mcteq}
\begin{multline}\label{mcttimedom}
  \bs\tau(q)\dot{\bs\Phi}(q,t)+\bs S(q)^{-1}\bs\Phi(q,t)\\
  +\int_0^t\bs M(q,t-t')\dot{\bs\Phi}(q,t')\,dt'=0\,.
\end{multline}
It is to be understood as a matrix equation as is indicated by the bold symbols.
$\bs S(q)$ is the matrix of partial structure factors defined by
$S_{\alpha\beta}(q)=\langle\varrho_\alpha(\vec q)|\varrho_\beta(\vec q)\rangle$.
The short-time behavior of the correlators
is given by $\bs\Phi(q,t)=\bs S(q)-\bs\tau(q)^{-1}|t|+{\mathcal O}(t^2)$,
where $\bs\tau(q)$ is a symmetric positive definite matrix of relaxation times.
It shall be specified in terms of short-time diffusion coefficients
$D^0_\alpha$ as $\tau_{\alpha\beta}(q)=1/(q^2D^0_\alpha)
\delta_{\alpha\beta}$.
The memory kernel $\bs M(q,t)$ is given through the so-called fluctuating
forces, $M_{\alpha\beta}(q,t)=(k_{\text{B}}T)^2(x_\alpha/m_\alpha)(x_\beta/
m_\beta)
\langle{\mathcal Q}{\mathcal L}j_{\alpha}(\vec q)|{\mathcal R}'(t)
{\mathcal Q}{\mathcal L}j_{\beta}(\vec q)\rangle$.
Here, $\mathcal Q$ is the projector perpendicular to the
number densities, $\varrho_\alpha(\vec q)$, and the longitudinal parts of the
number current densities, $qj_\alpha(\vec q)={\mathcal L}
\varrho_\alpha(\vec q)$. ${\mathcal R}'(t)=\exp[i{\mathcal Q}{\mathcal L}
{\mathcal Q}t]$ is the reduced evolution operator.
The $m_\alpha$ are the masses of the species labeled by $\alpha$, and
$x_\alpha=N_\alpha/N$ are the number concentrations.

Equation~(\ref{mcttimedom}) is complemented by an approximate expression for the
memory kernel. The MCT approximation for this quantity follows
from a straightforward generalization of the one-component case
\cite{Goetze1987b} and gives the polar form
\begin{equation}\label{mctfunctional}
 \bs M(q,t)={\mathcal F}[\bs\Phi(t),\bs\Phi(t)](q)
\end{equation}
of a symmetric bilinear form of the density correlators
\begin{widetext}
\begin{equation}\label{mctmemory}
  {\mathcal F}_{\alpha\beta}[\bs\Phi^{(1)},\bs\Phi^{(2)}](q)=
  \frac1{2q^2}\frac{\varrho}{x_\alpha x_\beta}
  \sum_{\alpha'\beta'\alpha''\beta''}\sum_{\vec k}
  V_{\alpha\alpha'\alpha''}(\vec q,\vec k,\vec p)
  \Phi_{\alpha'\beta'}^{(1)}(k)\Phi_{\alpha''\beta''}^{(2)}(p)
  V_{\beta\beta'\beta''}(\vec q,\vec k,\vec p)\,.
\end{equation}
\end{widetext}
Here, $\rho$ is the total number density, $\vec p=\vec q-\vec k$, and
$V_{\alpha\beta\gamma}(\vec q,\vec k,\vec p)$ are vertices quantifying the
coupling of a force fluctuation of wave vector $\vec q$ to density-fluctuation
pairs with wave vectors $\vec k$ and $\vec p$, respectively. The vertices are
given by the equilibrium structure of the system in terms of the
Ornstein-Zernike direct correlation function $c_{\alpha\beta}(q)$ and static
three-particle correlations. The latter shall be expressed in terms of
the $S_{\alpha\beta}(q)$ using the convolution approximation.
One thus arrives at
\begin{equation}\label{mctvertex}
  V_{\alpha\alpha'\alpha''}(\vec q,\vec k,\vec p)
  = (\vec q\vec k/q)c_{\alpha\alpha'}(k)\delta_{\alpha\alpha''}
  +(\vec q\vec p/q)c_{\alpha\alpha''}(p)\delta_{\alpha\alpha'}\,.
\end{equation}
\end{subequations}

A few remarks to these equations might be in order. The solution
to Eqs.~(\ref{mcteq}) exists for all $t\ge0$ and it is
uniquely determined by the initial conditions.
For a system with colloidal short-time dynamics,
the correlation functions $\bs\Phi(q,t)$ are completely monotone functions
\cite{Naegele1996}. This property is preserved by the specified MCT
approximation. In detail, it implies the following.
The matrices $\bs\Phi(q,t)$ are positive
definite, written $\bs\Phi(q,t)\succeq0$, for all times $t$ and at all $q$;
and for the time derivatives,
there holds $(-1)^l\partial_t^l\bs\Phi(q,t)\succeq0$ for
all $l=1,2,\ldots$.
Furthermore, the solution depends smoothly on $S_{\alpha\beta}(q)$,
$\tau_{\alpha\beta}(q)$ and $V_{\alpha\beta\gamma}(\vec q,\vec k,\vec p)$
for any fixed finite time interval. At
$t\to\infty$, bifurcations in the long-time limit
$\bs F(q)=\lim_{t\to\infty}\bs\Phi(q,t)$ may occur at critical points
called glass-transition singularities.
The $F_{\alpha\beta}(q)$ are called the glass form factors and they
are solutions of the equation
\begin{subequations}\label{mcteq_fandh}
\begin{equation}\label{mctfeq}
  \bs F(q)=\bs S(q)-\left[\bs S(q)^{-1}+{\mathcal F}[\bs F, \bs F](q)\right]
  ^{-1}\,.
\end{equation}
In particular, the correct $\bs F(q)$ can be determined through an iteration
scheme, $\bs F^{(n)}(q)=\bs S(q)-\left[\bs S(q)^{-1}+{\mathcal F}
[\bs F^{(n-1)},\bs F^{(n-1)}](q)\right]^{-1}$, $n=1,2,\ldots$,
with starting point $\bs F^{(0)}(q)=\bs S(q)$. The sequence $\bs F^{(n)}(q)$
converges monotonically towards $\bs F(q)$.
For sufficiently small vertices, one has the liquid solution
$\bs F(q)=\bs0$, while the glass is characterized by $\bs F(q)\neq\bs0$.
Solutions corresponding to critical points
shall be denoted by a superscript $c$.
To understand the bifurcation scenario, one needs to discuss the
critical eigenvector of the linearization of Eq.~(\ref{mctfeq}), $\bs H(q)$,
given through
\begin{equation}\label{mctheq}
  \bs H(q)-2(\bs S^c(q)-\bs F^c(q)){\mathcal F}^c[\bs F^c,\bs H](q)
  (\bs S^c(q)-\bs F^c(q))=\bs0\,.
\end{equation}
\end{subequations}
This eigenvector is nondegenerate, which implies that all MCT bifurcations
belong to the type $A_\ell$, $\ell=2,3,\ldots$, introduced by Arnol'd
\cite{Arnold1986}. Also, one can choose $\bs H(q)\succ0$.
All of the preceding remarks apply to the general case of matrix-valued MCT
equations; they are not affected by the precise form of the
vertices entering Eq.~(\ref{mctmemory}).
Proofs of the cited mathematical properties of Eqs.~(\ref{mcteq})
and (\ref{mcteq_fandh}) can be found in Ref.~\cite{Franosch2002}.
In this paper only the simplest bifurcations, i.e.\ those of type $A_2$
are discussed,
where a jump occurs in $\bs F(q)$ from $\bs0$ to the critical
value $\bs F^c(q)\succ0$.

There is an important implication of the cited general results that
can be substantiated in complete analogy to the one obtained for
one-component systems \cite{Franosch1998,Fuchs1999b}.
There exists a time scale $t_0$ such that one can express the solution
of Eq.~(\ref{mcteq}) for $t\gg t_0$ in the form
\begin{subequations}\label{mctstruc}
\begin{equation}\label{mctstrucdef}
  \bs\Phi(q,t)=\bs S^{1/2}(q)\bs\phi(q,\tilde t)\bs S^{1/2}(q)\,,
\end{equation}
where $\tilde t=t/t_0$.
The time scale $t_0$ depends smoothly on $S_{\alpha\beta}(q)$,
$\tau_{\alpha\beta}(q)$, and $V_{\alpha\beta\gamma}(\vec q,\vec k,\vec p)$.
The sensitive dependence of $\bs\Phi$ on the control parameters is
described by the completely monotone function $\bs\phi(q,\tilde t)$.
The latter is determined, up to some arbitrary time scale $t_*$, through
the equation
\begin{equation}\label{mctstruceq} \bs\phi(q,t)=\bs m(q,t)
  -\frac{d}{dt}\int_0^t\bs m(q,t-t')\bs\phi(q,t')\,dt'\,,
\end{equation}
obeying the initial condition
\begin{equation}\label{mctstrucin}
  \lim_{t\to0}(t/t_*)^{1/3}\bs\phi(q,t)=\bs 1\,.
\end{equation}
Here the kernel $\bs m(q,t)$ is given by the mode-coupling functional,
Eq.~(\ref{mctmemory}), as
\begin{multline}\label{mctstrucmem}
  \bs m(q,t)=\bs S^{1/2}(q)\times\\ \times
  {\mathcal F}
  \left[\bs S^{1/2}\bs\phi(t)\bs S^{1/2},\bs S^{1/2}\bs\phi(t)\bs S^{1/2}
  \right](q)
  \bs S^{1/2}(q)\,.
\end{multline}
\end{subequations}
This means that $\bs\phi(q,\tilde t)$ is determined by the quantities
entering $\mathcal F$ only, i.e.\ by the parameters specifying the equilibrium
structure. In this manner, MCT justifies the concept of structural relaxation
as opposed to, e.g., transient relaxation. The latter, be it Brownian or
Newtonian, merely enters the scale $t_0$. In particular this implies that
details of $\tau_{\alpha\beta}(q)$ do not affect the structural relaxation
apart from influencing $t_0$.

Let us also note the formulas for the longitudinal and transversal elastic
moduli of the mixture. They are given through Green-Kubo relations involving
the total mass currents \cite{Mryglod1997,
Naegele1998}. One defines the projector
${\mathcal Q}_{\text{HD}}$ as projecting out $\varrho_\alpha(\vec q)$ and the
longitudinal mass current, $J(\vec q)=\sum_\alpha m_\alpha j_\alpha(\vec q)$,
together with the corresponding reduced resolvent
${\mathcal R}'_{\text{HD}}(z)$.
The latter is the Laplace transform for frequency $z$ of the corresponding
evolution operator ${\mathcal R}'_{\text{HD}}(t)=\exp [i{\mathcal Q}_{\text{HD}}
{\mathcal L}{\mathcal Q}_{\text{HD}}t]$.
For the longitudinal viscosity, this yields
\cite{Mryglod1997}
\begin{equation}\label{etal} \eta^{\text{L}}
  =\lim_{z\to0}\lim_{q\to0}\frac{\varrho}{k_{\text{B}}T}\frac1{q^2}
  \langle{\mathcal Q}_{\text{HD}}{\mathcal L}J(\vec q)|
  {\mathcal R}'_{\text{HD}}(z){\mathcal Q}_{\text{HD}}{\mathcal L}J(\vec q)
  \rangle\,.
\end{equation}
At the bifurcation singularity, a nontrivial long-time
limit $\bs F^c(q)$ implies a pole at zero frequency,
$\bs\Phi^c(q,z)\sim-\bs F^c(q)/z$.
According to this, the longitudinal modulus shows a discontinuity
$\delta M_{\text{L}}^c$ at the glass transition. Rewriting
Eq.~(\ref{etal}) in terms of the MCT projector $\mathcal Q$, one gets
\begin{equation}\label{modl}
  \delta M^c_{\text{L}}=(\varrho k_{\text{B}}T)
  \lim_{q\to0}\sum_{\alpha\beta}x_\alpha
  {\mathcal F}^c_{\alpha\beta}[\bs F^c,\bs F^c] (q) x_\beta\,.
\end{equation}
The glass, different from the liquid, is characterized by a finite
shear modulus. A formula
similar to Eq.~(\ref{modl}) is obtained for the
shear modulus $M_{\text{T}}^c$ at the glass transition,
\begin{equation}\label{modt}
  M^c_{\text{T}}=(\varrho k_{\text{B}}T)
  \lim_{q\to0}\sum_{\alpha\beta}x_\alpha
  {\mathcal F}^{\text{T},c}_{\alpha\beta}[\bs F^c,\bs F^c](q)
  x_\beta\,.
\end{equation}
Here, the MCT expression for the transverse
fluctuating-force kernel ${\mathcal F}^{\text{T}}(q)$ is obtained
from ${\mathcal F}(q)$ by replacing in Eq.~(\ref{mctmemory}) the vertices by
\begin{multline}\label{vertext}
  {\mathcal V}^{\text{T}}_{\alpha\alpha'\alpha''}(\vec q,\vec k,\vec p)
  = (\vec q^{\text{T}}\vec k/q)
     c_{\alpha\alpha'}(k)\delta_{\alpha\alpha''}\\
  + (\vec q^{\text{T}}\vec p/q)
     c_{\alpha\alpha''}(p)\delta_{\alpha\alpha'}\,.
\end{multline}
In this formula, $\vec q^{\text{T}}$ is a vector of length $q$ perpendicular
to $\vec q$.

\subsection{The Binary Hard-Sphere Mixture}

The general theory shall be applied to binary hard-sphere mixtures (HSM),
consisting of large ($\text{A}$) and small ($\text{B}$) particles.
If $d_\alpha$, $\alpha=\text{A},\text{B}$, denote the particle diameters,
the packing fractions of the species read $\varphi_\alpha=(\pi/6)(x_\alpha
\varrho)d_\alpha^3$, and the total packing fraction is given by
$\varphi=\varphi_{\text{A}}+\varphi_{\text{B}}$. The thermodynamic state
is characterized by three control parameters.
Let us choose them to be the total packing fraction $\varphi$, the size ratio
$\delta=d_{\text{B}}/d_{\text{A}}\le1$,
and the packing contribution of the smaller
species $\hat x_{\text{B}}=\varphi_{\text{B}}/\varphi$.
Whenever composition changes are considered in the
following, a variation of $\hat x_{\text{B}}$ for fixed $\varphi$
and $\delta$ is to be understood.
This in turn implies the number concentration of the small particles to vary
as
\begin{equation}\label{numberconc}
  x_{\text{B}}=\frac{\hat x_{\text{B}}/\delta^3}
                    {1+\hat x_{\text{B}}(1/\delta^3-1)}\,.
\end{equation}
The procedure is somewhat in between a true addition, which would increase
both total density and total packing fraction, and a replacement of a certain
amount of large spheres by the same amount of smaller ones, which would reduce
the total packing fraction.
For sufficiently small $\delta$, there appears a percolation threshold
for the motion of the small particles in the glass formed by the large
ones. This transition and its precursor phenomena shall not be considered in
this paper.

Static structure input for our model is taken from the Percus-Yevick (PY)
approximation \cite{Lebowitz1964b,Baxter1970}.
More accurate solutions of the Ornstein-Zernike integral equations for
hard-sphere mixtures are
available. Yet, one knows from the one-component MCT that improvements
aiming at, for example,
thermodynamic consistency have little influence on the glassy
dynamics.
Unfortunately, the quality of the PY approximation at the desired high packing
fractions
is unknown. It is known that large errors of the structure factor can occur if
one goes
over to large values of $1/\delta$ \cite{Malijevsky1997}, but this case is
excluded from our discussion.

With the structure factor and the direct correlation functions given, the
vertices in Eqs.~(\ref{mctvertex}) and (\ref{vertext}) are well defined
functions of the wave vectors and matrix indices. Hence, also the
mode-coupling functional $\mathcal F$ in Eq.~(\ref{mctmemory}) is defined as
a triple integral over the components of $\vec k$; and the same holds for
the functional ${\mathcal F}^T$. After introduction of bipolar coordinates
and using rotational symmetry,
the $\vec k$ integrals are transformed to double integrals over $k=|\vec k|$
and $p=|\vec q-\vec k|$. After performing the $\vec q\to0$ limit in the
functionals, the zero-wave-vector limits entering Eqs.~(\ref{modl}) and
(\ref{modt}) are reduced to one-dimensional integrals over $k$.
As a next step, the wave vectors are reduced to points on a grid of $M$
values. The grid is chosen as $qd_{\text{A}}=q_0+\hat q\Delta q$, with
$q_0=0.2$, $\Delta q=0.4$, and $\hat q=0,1,\ldots,M-1$, unless otherwise
stated. The integrals are replaced by Riemann sums. The resulting formulas
are the same as explained explicitly before for the one-component systems
\cite{Franosch1997}, but additional sums over matrix indices occur.
As a result, the cited equations refer to ones for sets of $M$ matrix
correlators, where $q$ serves as a label for the correlators. To complete
the specification of the equations, the short-time diffusion constants
are taken according to Stokes' law, $D^0_\alpha=C/d_\alpha$. The unit of
time is chosen so that $C=0.01$.

For the simple hard-sphere system, it was found that a choice of $M=100$
is sufficient to avoid cut-off and discretization effects for the
results \cite{Franosch1997}.
Our work was done mostly with $M=200$, implying a cut-off wave
vector $q^*d_{\text{A}}=79.8$. This enables us to handle size ratios
$\delta\ge0.5$ with the accuracy used earlier for the simple system. Thus,
Eqs.~(\ref{mcttimedom})--(\ref{mctmemory}) are 600 coupled integro-differential
equations for 600 correlators, and Eq.~(\ref{mctfeq}) formulate 600 implicit
equations for the 600 glass form factors. The latter equations are
solved by the iteration mentioned above. Thereby, one gets the form factors,
discussed in Sec.~\ref{sec.nep}.
Shifting the value for the packing fraction $\varphi$, one identifies
the glass transition points.
The linear equation for the nondegenerate eigenvector $\bs H(q)$,
Eq.~(\ref{mctheq}), is solved by a standard routine. This eigenvector is
used to calculate the asymptotic solutions discussed in Sec.~\ref{sec.dynplat}.
The critical glass form factors are substituted into Eqs.~(\ref{modl})
and (\ref{modt}) so that the contributions to the moduli discussed
in Sec.~\ref{sec.phase} can be calculated.

The closed set of Eqs.~(\ref{mcttimedom})--(\ref{mctmemory}) as well as
Eqs.(\ref{mctstrucdef})--(\ref{mctstrucmem}) for the time dependence of the
correlators are solved by a method adapted to this special kind of
Volterra problem. The solutions of Eqs.~(\ref{mcteq}) yield the correlators
to be compared with the cited light scattering data. The solutions of
Eqs.~(\ref{mctstruceq}) are used in Sec.~\ref{sec.dynplat} to separate
structural relaxation from the transient dynamics. To proceed, one
introduces a grid on the time axis of equal step size $h$ consisting of
$N$ points. The time derivatives in Eqs.~(\ref{mcttimedom}) and
(\ref{mctstruceq}) are replaced by difference relations and the
convolution integrals by Riemann sums. One gets a recursion relation
determining the solutions for time $t_{n+1}=h(n+1)$ from the values
$t_l$, $l\le n$. The initial values are taken from the short-time asymptote,
given by $\bs\Phi(q,t)-\bs S(q)\sim-\bs\tau(q)^{-1}t$ or Eq.~(\ref{mctstrucin}),
respectively. Because of the scale invariance of the equations of structural
relaxation, the value of $t_*$ does not matter. One has to make sure that the
results obtained for $0<t<t_N$ remain stable within the desired accuracy
upon doubling $h$ or halving $N$. Then one carries out a decimation by
setting $h\mapsto 2h$. Thereby, one solves the equations up to $2t_N$. This
procedure is repeated until the correlators reach their long-time asymptote.
Details are explained, e.g., in Ref.~\cite{Goetze1996}. Typically,
the figures in this paper have been calculated with $h=10^{-6}$ and
$N=256$. Let us point out that Eqs.~(\ref{mcteq})--(\ref{vertext})
are completely analogous to the ones discussed repeatedly in previous studies
of one-component systems. Also the numerical methods applied for the solution
are the ones used earlier for the simple case. The additional complication
here is the handling of matrices; but this is straightforward though
numerically more demanding.

\section{Transition Diagram}\label{sec.phase}

\begin{figure}
\includegraphics[width=\figwidth]{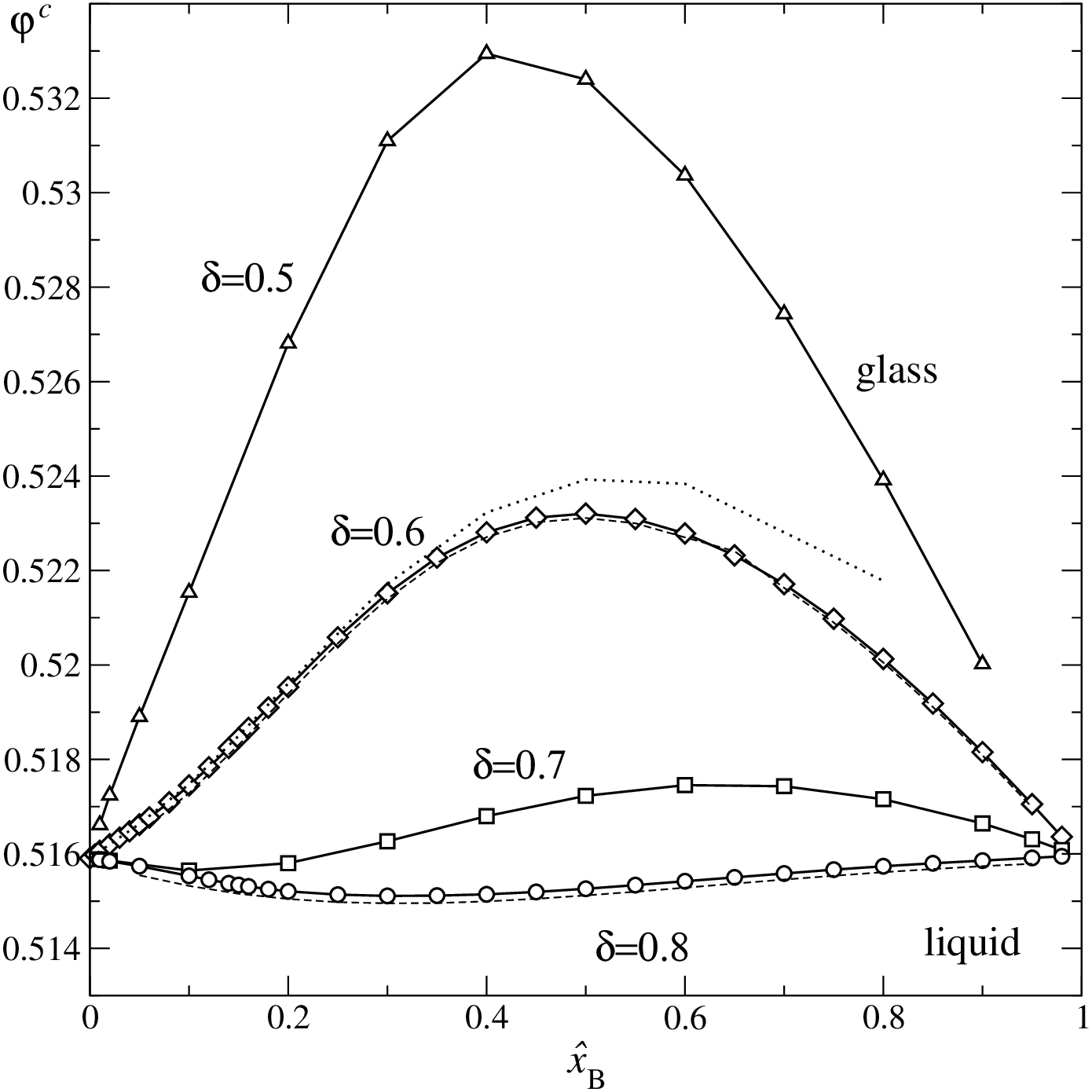}
\caption{\label{phasediag}
  Liquid-glass transition diagram of a binary hard sphere mixture (HSM) for
  size ratios $\delta=0.5$ (triangles),
  $\delta=0.6$ (diamonds), $\delta=0.7$ (squares), and
  $\delta=0.8$ (circles), plotted as critical total packing fraction
  $\varphi^c$ versus packing contribution of the smaller
  species, $\hat x_{\text{B}}=\varphi_{\text{B}}/\varphi$.
  Full lines are guides to the eyes. The dashed lines indicate results
  obtained from tripling the number $M$ of grid points from $M=200$ to $M=600$,
  and the dotted line for $\delta=0.6$ shows results obtained using $M=100$;
  see text for details.
}
\end{figure}

Cuts through the liquid-glass transition surface in the three-dimensional
control-parameter space for the binary HSM are
depicted in Fig.~\ref{phasediag}.
To assure that the results do not seriously depend on the discretization
used, we show as well the glass transition points calculated for
$\delta=0.6$ and $0.8$ from
the model with $M=600$, $\Delta q=0.4/3$, $q_0=0.2/3$. In addition,
for $\delta=0.6$ the dotted line exhibits the result calculated
with cutoff $q^*d_{\text{A}}=39.8$ and $M=100$
which are the discretization parameters used in Ref.~\cite{Franosch1997}.
One infers that for $\hat x_{\text{B}}\lesssim0.3$, this discretization
would be sufficient to produce reasonable results.

For fixed size ratio $\delta\lesssim0.65$, the critical packing fraction
first increases upon increasing $\hat x_{\text{B}}$. Since
$\hat x_{\text{B}}=0$ and $\hat x_{\text{B}}=1$ both represent monodisperse
hard-sphere systems,
one gets $\varphi^c(\hat x_{\text{B}}=0)=\varphi^c(\hat x_{\text{B}}=1)$.
Thus, the liquid-glass transition lines for $\delta\lesssim0.65$
exhibit a maximum at some intermediate values of $\hat x_{\text{B}}$.
It is well understood that big particles moving in a liquid of much smaller
ones experience an effective attraction \cite{Asakura1958} that is of purely
entropic origin. Such a short-ranged attraction leads to a stabilization of
the liquid phase, as was explained earlier
\cite{Fabbian1999,Bergenholtz1999,Dawson2001}. Our result is a direct
analogon of this depletion-attraction
effect. Similarly, from the discussion of polymer
melts it is known that the introduction of smaller components into the
system typically decreases the viscosity, i.e., drives the system further into
the liquid phase; an effect sometimes called ``plasticizing''.
Therefore, the effect found here is an entropically
induced plasticization effect.

For less-disparate-sized mixtures, the theory predicts an inversion of the
effect described above. An example is shown in Fig.~\ref{phasediag} for
$\delta=0.8$, where a decrease of $\varphi^c$ with increasing
$\hat x_{\text{B}}$
up to some minimum point is observed. This is in accordance with similar
MCT results for a binary soft-sphere mixture \cite{Barrat1990b}.
It means that the introduction of disorder due to a small polydispersity
of the particles stabilizes the glass state.
The transition diagram is not symmetric
with respect to $\hat x_{\text{B}}\to(1-\hat x_{\text{B}})$; our theory predicts
for $0.65\lesssim\delta\lesssim0.8$ ``S''-shaped transition lines.

\begin{figure}
\includegraphics[width=\figwidth]{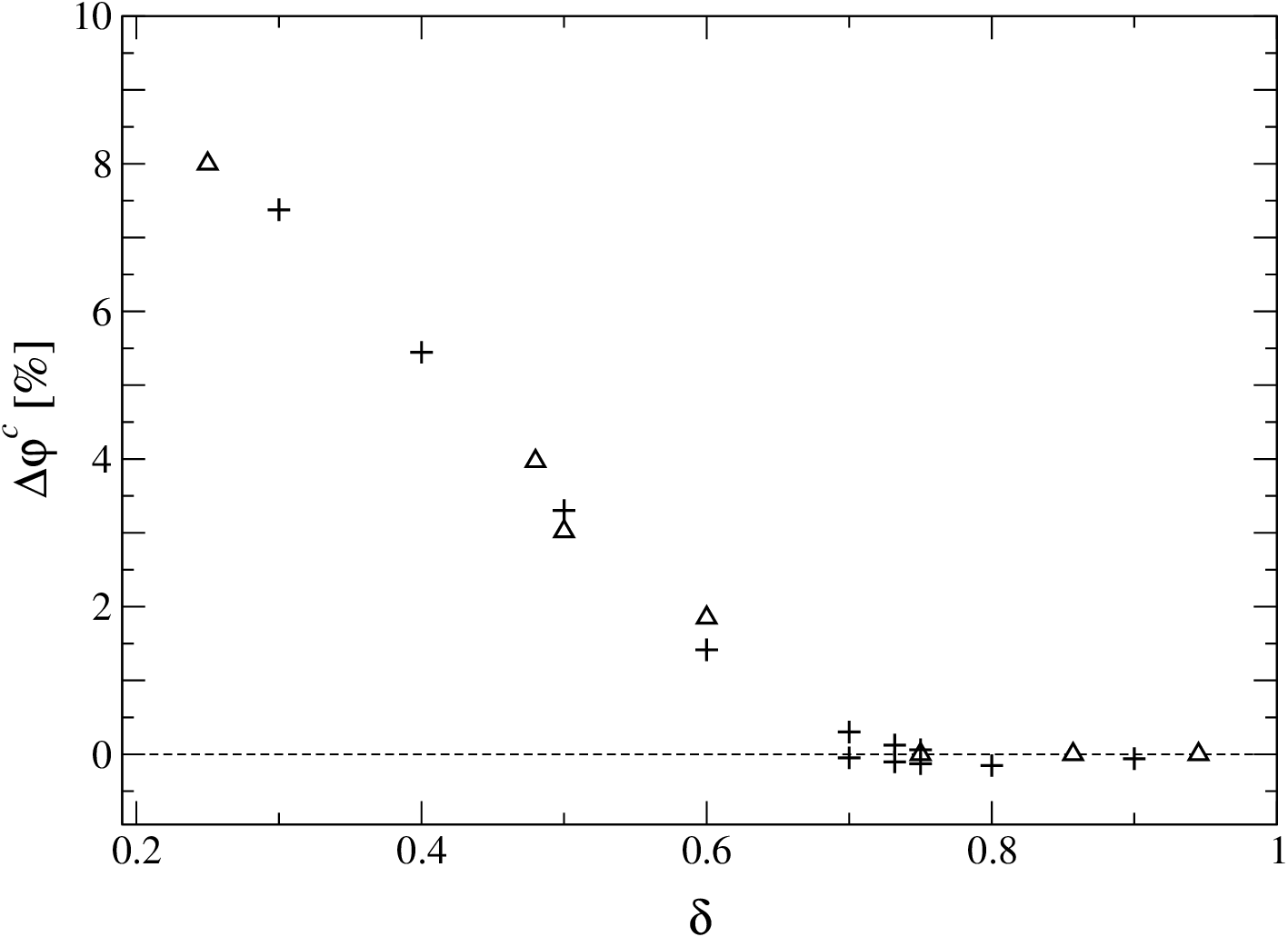}
\caption{\label{relincr}
  Maximum relative increase and decrease of the critical packing fraction,
  $\Delta\varphi^c(\delta)$, according to Eq.~(\ref{phirelative}), as a function
  of the size ratio $\delta$
  (crosses), together with experimental data for random loose packing
  (triangles, reproduced from Ref.~\protect\cite{Lemaignan1980}, cf.\ text).
  For the MCT critical packing fraction values, two symbols are noted for
  those $\delta$, where a maximum and a minimum different from the
  $\delta=1$ value could be identified.
}
\end{figure}

To get another view on the transition diagram, let us define the relative change
of $\varphi^c$ with respect to the one-component case through
\begin{equation}\label{phirelative}
  \Delta\varphi^c(\delta)=(\varphi^c(\delta,x_{\text{B}}^\pm)-\varphi_0^c)/
  \varphi_0^c \,.
\end{equation}
Here,
$x_{\text{B}}^\pm$ are the points at which a maximum or a minimum
occurs in $\varphi^c(x_{\text{B}})$ for fixed $\delta$; $\varphi_0^c
\approx0.5159$
is the critical packing fraction of the one-component system.
The resulting values are plotted in Fig.~\ref{relincr} together with
data taken from Ref.~\cite{Lemaignan1980}. There, results for
$\Delta\varphi(\delta)$
of several experiments for random-loose
sphere packings in two-component steel-ball mixtures have been presented.
These are
operationally defined as the random packing fractions obtained when pouring
spheres into a container without subsequent densification through
shaking.
One observes that both $\Delta\varphi^c(\delta)$ and the data follow the same
trend.
Note that we get negative values
for $0.65\lesssim\delta<1$. In Ref.~\cite{Lemaignan1980}, no such effect is
discussed, but it is reported that there seems to be no observable change in
the data.
There is no precise theoretical definition of the concept of random loose
packing. Nevertheless, the
reported values can be taken as a quantization of
a mixing effect,
i.e.\ of modifications of the random cage structure.
The fact that the variation in $\Delta\varphi^c$ with $\delta$ agrees with
these experimental findings supports the conclusion that MCT is able to
capture the change in the average cage structure induced by the presence
of the second component.

\begin{figure}
\includegraphics[width=\figwidth]{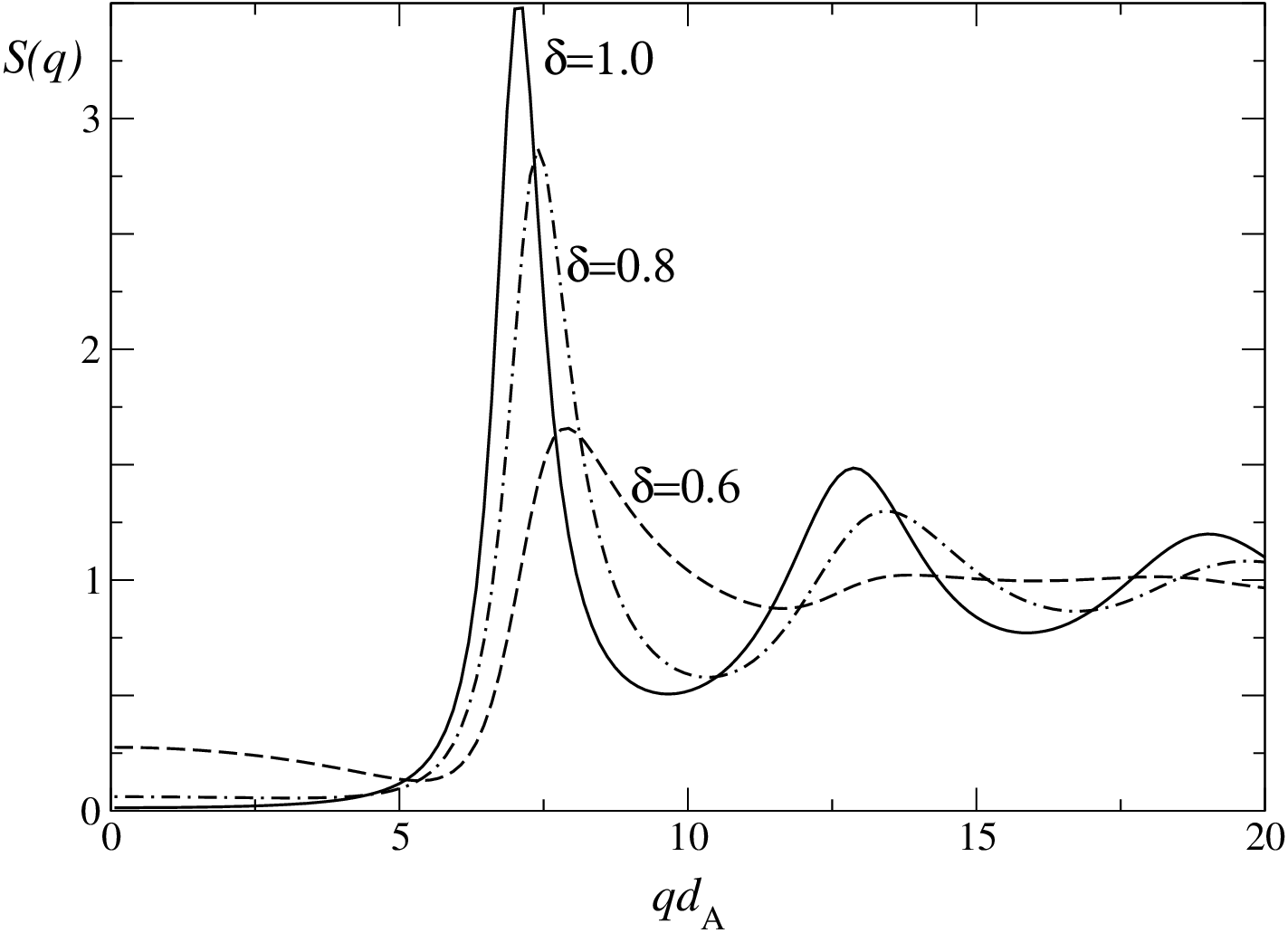}
\caption{\label{sqtotal}
  Total structure factor, $S(q)=\sum_{\alpha\beta}S_{\alpha\beta}(q)$,
  calculated in Percus-Yevick approximation for binary mixtures with
  $\varphi=0.515$, $\hat x_{\text{B}}=0.2$ and three values of $\delta$.
}
\end{figure}

The results from above suggest that the change of the glass-transition point
with composition can be understood by looking at the geometrical
structure of the system. This information is reflected by the static
structure factors, which comprise the relevant input for the MCT vertex in
Eq.~(\ref{mctvertex}). In particular, it is understood that the $q$-vector
region around the first sharp peak in $S(q)$ is important for explaining the
MCT glass transition. Fig.~\ref{sqtotal} shows this region for the total
structure factor, $S(q)=\sum_{\alpha\beta}S_{\alpha\beta}(q)$,
at fixed packing fraction $\varphi=0.515$, composition
$\hat x_{\text{B}}=0.2$, and different $\delta$. One notices two trends
caused by decreasing $\delta$,
viz.\ a decrease in peak height and an increase in its large-$q$ wing.
The interplay between these two trends is responsible for the shift in
$\varphi^c$. At larger $\delta$, the increase in the
wing is dominant and stabilizes the glass, i.e.\ it reduces $\varphi^c$ with
respect to the one-component
system. But at $\delta\lesssim0.65$, the reduction in peak height, equivalent
to a weakening of the intermediate-range order, overwhelms this trend.
This effect stabilizes the liquid, i.e.\ increases $\varphi^c$.
In all cases, the peak position shifts to higher $q$, indicating that, on
average, particles are closer together in the mixture than in the one-component
system; an effect typical for the introduction of effective attractive
interactions \cite{Dawson2001}.

\begin{figure}
\includegraphics[width=\figwidth]{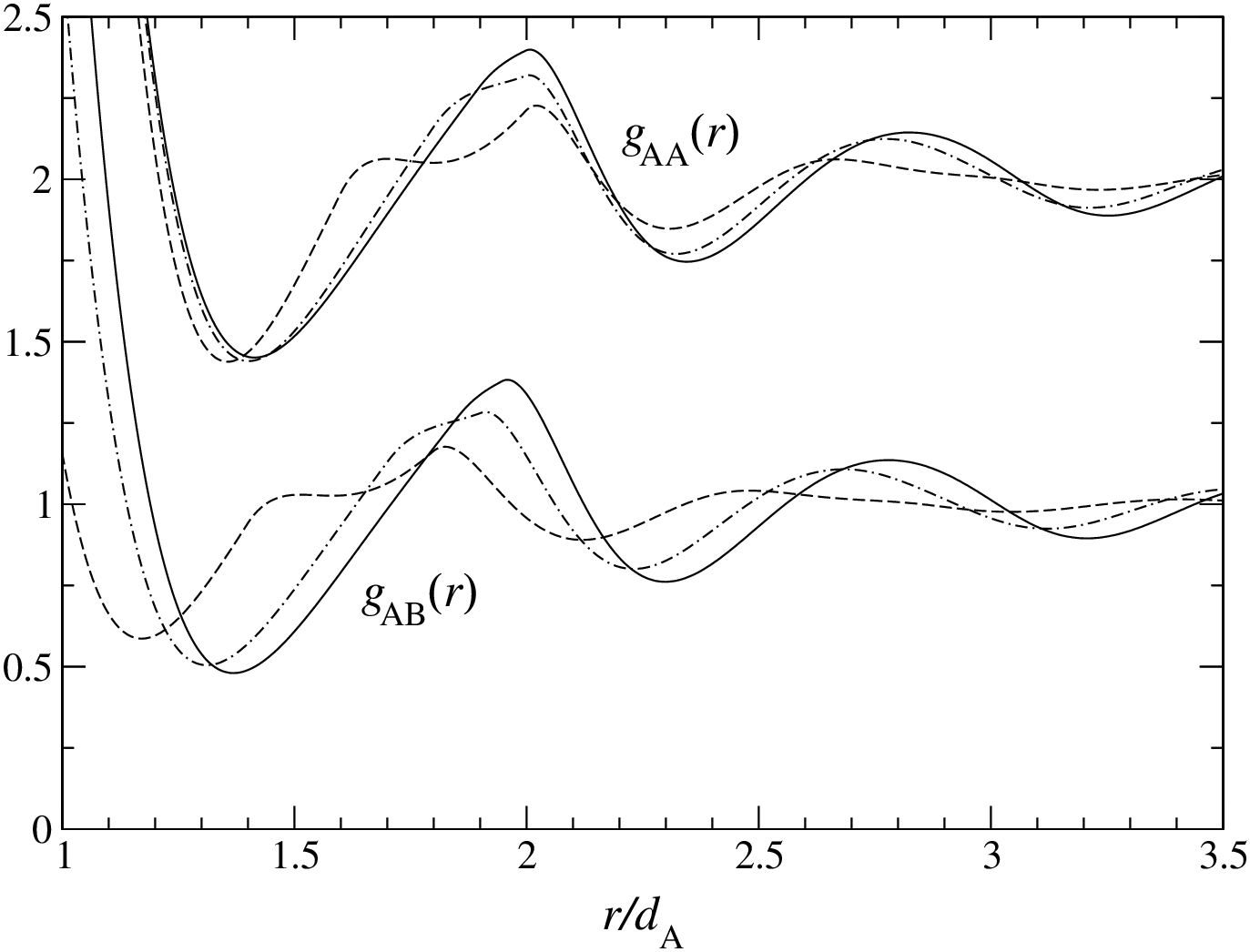}
\caption{\label{pairdistrib}
  Results within the Percus-Yevick approximation
  for the partial pair correlation functions
  $g_{\text{AA}}(r)$ and $g_{\text{AB}}(r)$ of binary HSM
  at $\varphi=0.516$, $\hat x_{\text{B}}=0.2$, and
  $\delta=0.9$ (solid lines), $\delta=0.8$ (dot-dashed lines), and
  $\delta=0.6$ (dashed lines). Curves for $g_{\text{AA}}(r)$ have been shifted
  up by $1.0$ for clarity.
}
\end{figure}

Another way of looking at the local structure of the HSM
is provided by the partial pair distribution functions, $g_{\alpha\beta}(r)$.
These have been obtained by numerically solving the Ornstein-Zernike
equation in the $r$-domain using Baxter's factor function for the PY closure.
Results are
shown in Fig.~\ref{pairdistrib} for $g_{\text{AA}}(r)$ and $g_{\text{AB}}(r)$,
again at fixed $\varphi$ and $\hat x_{\text{B}}$ for various $\delta$.
Here, both quantities vary more or less in phase for $\delta\gtrsim0.7$,
indicating that the local ordering of the one-component system is only
slightly disturbed.
One infers the $\text{B}$ particles to be responsible for smaller
average particle distances, thus favoring
arrest of the structure. For smaller $\delta$, the one-component system's
structure is modified more severely,
since $g_{\text{AA}}(r)$ and $g_{\text{AB}}(r)$
no longer vary in phase. Instead, ``chemical ordering'' effects can be
seen, and they are responsible for the shift of the glass transition to higher
packing fractions.

Let us stress that the variation of $\varphi^c$ with concentration, while
being small in total, nevertheless has a large impact on the dynamics close
to the glass transition. This holds since relaxation times of the liquid
in this region depend strongly on the distance to the critical
packing fraction. We shall recur to this point in
Sec.~\ref{sec.dyn}A.

\section{Glass Form Factors}\label{sec.nep}

The spontaneous arrest of density fluctuations within the glass state
is quantified by the glass form factors $F_{\alpha\beta}(q)$. In
principle, these quantities can be measured in a scattering experiment
via the intensity of the elastic line in the cross section. The diagonal
elements $\hat f_{\alpha\alpha}(q)$ of the normalized quantities
\begin{equation} \hat f_{\alpha\beta}(q)=F_{\alpha\beta}(q)
  /\sqrt{S_{\alpha\alpha}(q)S_{\beta\beta}(q)}
\end{equation}
have the meaning of the Debye-Waller factor for the distribution of
species $\alpha$. In the limit $\hat x_{\text{B}}\to0$,
$\hat f_{\text{BB}}(q)$
is the spatial Fourier transform of the density distribution of a single
localized $\text{B}$
particle. It is the Lamb-M\"o{\ss}bauer factor $f^s_{\text{B}}(q)$ of a
$\text{B}$ particle in the hard-sphere system of $\text{A}$ particles.
A similar statement holds with the role of $\text{A}$ and $\text{B}$
particles interchanged, i.e., $f^s_{\text{A}}(q)=\hat f_{\text{AA}}(q,
\hat x_{\text{B}}\to1)$, but then the tagged particle is of the size
$1/\delta$ in units of the surrounding hard spheres' diameter.
If the packing fraction $\varphi$ decreases towards the transition value
$\varphi^c$, the $\hat f_{\alpha\alpha}(q)$ decrease towards their
critical values, $\hat f_{\alpha\alpha}^c(q)$. These values are of particular
relevance since they specify the so-called plateau values of the correlation
functions of the liquid for states near the liquid-glass transition
\cite{Goetze1991b}.
This will be discussed further in Sec.~\ref{sec.dyn}.

\begin{figure}
\includegraphics[width=\figwidth]{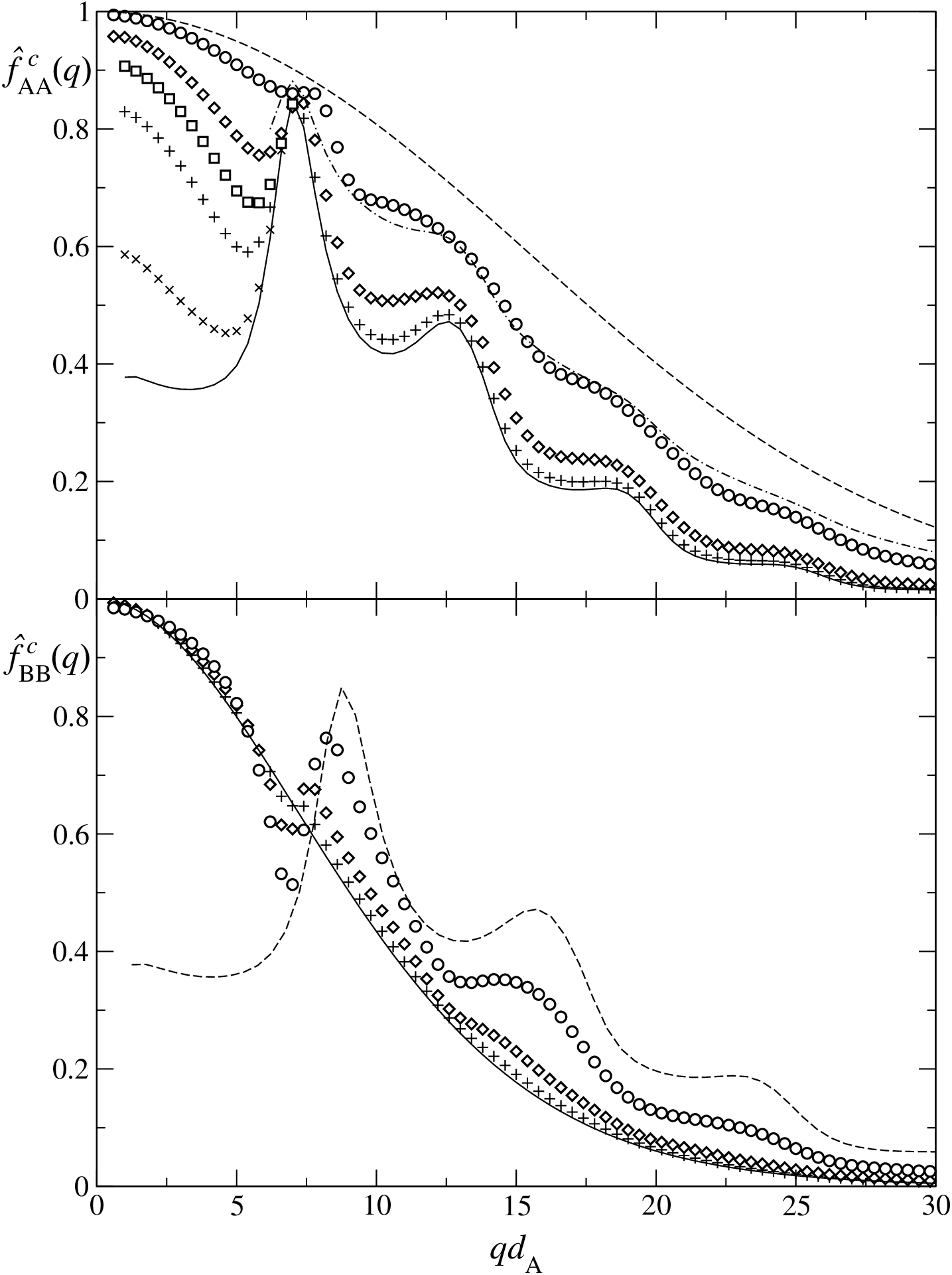}
\caption{\label{fig-fc08}
  Critical glass form factors $\hat f^c_{\alpha\alpha}(q)
  =F^c_{\alpha\alpha}(q)/S_{\alpha\alpha}(q)$ of a binary HSM with
  size ratio $\delta=0.8$ for the large
  (upper panel) and small (lower panel) particles.
  The packing contributions of the small spheres are $\hat x_{\text{B}}=0$
  (solid lines), $0.05$ (plus symbols),
  $0.2$ (diamonds), $0.6$ (circles), and $1.0$ (dashed lines).
  In the upper panel, results for small $q$ at $\hat x_{\text{B}}=0.01$
  (crosses) and $0.1$ (squares) are also shown.
  The dash-dotted line in the upper panel demonstrates the linear interpolation
  between the cases $\hat x_{\text{B}}=0$ and $\hat x_{\text{B}}=1$ for
  $qd_{\text{A}}\gtrsim6$.
}
\end{figure}

Fig.~\ref{fig-fc08} shows the critical Debye-Waller factors for small size
disparity, $\delta=0.8$, and various $\hat x_{\text{B}}$. One notices an
increase of the values with increasing $\hat x_{\text{B}}$ for
almost all $q$.
This result can be understood as follows.
With no second species present, $\hat f_{\text{AA}}^c(q)$ matches the
Debye-Waller factor of the one-component system, $f^c(q)$, shown by the full
line in the upper panel. On the other hand,
as mentioned above, for
$\hat x_{\text{B}}\to1$,
the quantity $\hat f_{\text{AA}}^c(q)$ crosses over to the
tagged-particle quantity of a bigger sphere in a surrounding fluid
of smaller ones, $f_{\text{A}}^{s,c}(q)$.
At $q\to0$ and $\hat x_{\text{B}}=0$, momentum conservation implies
$\hat f_{\text{AA}}^c(q\to0)<1$, while for $\hat x_{\text{B}}\to1$,
particle conservation
and momentum relaxation for the tagged particle require
$f_{\text{A}}^{s,c}(q\to0)=1$
\cite{Goetze1991b}. By interpolation, one gets an increase in
$\hat f^c_{\text{AA}}(q)$
with increasing $\hat x_{\text{B}}$ at small $q$.
For large $q$, on the other hand, the Debye-Waller
factor in a one-component system is oscillating
around the Lamb-M\"o{\ss}bauer factor of a tagged particle with equal
diameter.
The Lamb-M\"o{\ss}bauer factor in turn can be approximated reasonably
by a Gaussian,
$f^s(q)=\exp[-(qr_s)^2]$, where $r_s$ is the particle's
localization length \cite{Fuchs1998}.
The localization length becomes the smaller the bigger the radius $d^s$ of the
tagged particle is with respect to the radius $d$ of the surrounding spheres
\cite{Fuchs1998}; in particular one gets for a tagged particle of diameter
$d^s/d=1/0.6$ ($1/0.8$, $0.8$, $0.6$) the value $r^c_s/d=0.041$ ($0.056$,
$0.095$, $0.136$).
This implies the distribution of the $\hat f_{\text{AA}}^c(q)$,
given in the limit $\hat x_{\text{B}}\to1$ by $f^{s,c}(q)$ with
$d^s/d=\delta$, to be broader than that in the limit $\hat x_{\text{B}}\to0$,
given by $f^c(q)$.
Therefore the width of the distribution
$\hat f^c_{\text{AA}}(q)$ has to increase for $\delta<1$ as $\hat x_{\text{B}}$
increases from zero to unity. This is demonstrated in the upper panel
by the dash-dotted line. It represents a simple
interpolation, $\hat f^c_{\text{AA}}(q)\approx f^c(q)+(f^{s,c}_{\text{A}}(q)
-f^c(q))\hat x_{\text{B}}$ for $\hat x_{\text{B}}=0.6$ and
$q>6/d_{\text{A}}$.

The change of $\hat f^c_{\text{BB}}(q)$ can be understood along the same
line of reasoning. But one has to notice that in this case the localization
length of a smaller sphere in a surrounding of big ones matters.
In particular, one has $\hat f_{\text{BB}}^c(q,\hat x_{\text{B}}\to0)
=f_{\text{B}}^{s,c}(q/\delta)$. This yields a width of this distribution
smaller than the one of the $\hat f_{\text{BB}}^c(q,\hat x_{\text{B}}\to1)
=f^c(q/\delta)$.
Such an effect can be seen in the lower panel
of Fig.~\ref{fig-fc08} for $qd_{\text{A}}\gtrsim7$.
The crossover is naturally given by the size of the $\text{A}$ particles.
Based on the above argument,
one expects at smaller $q$ the inverse trend. But this is only found in
$\hat f^c_{\text{BB}}(q)$ for $5\lesssim qd_{\text{A}}\lesssim7$.
Instead one notices that for all $\hat x_{\text{B}}\le0.6$, the
$\hat f^c_{\text{BB}}(q)$ follow closely the result for $\hat x_{\text{B}}=0$,
i.e.\ they are still close to unity at small $q$. This is a consequence of
the normalization chosen here, since it is dominated by a change
in $S_{\text{BB}}(q)$ at small $q$. It could be eliminated when discussing
e.g.\ matrix-normalized quantities, $\bs f(q)=\bs S^{-1/2}(q)\bs F(q)
\bs S^{-1/2}(q)$, where the normalization properly accounts for the overall
change in $\bs S(q)$.

\begin{figure}
\includegraphics[width=\figwidth]{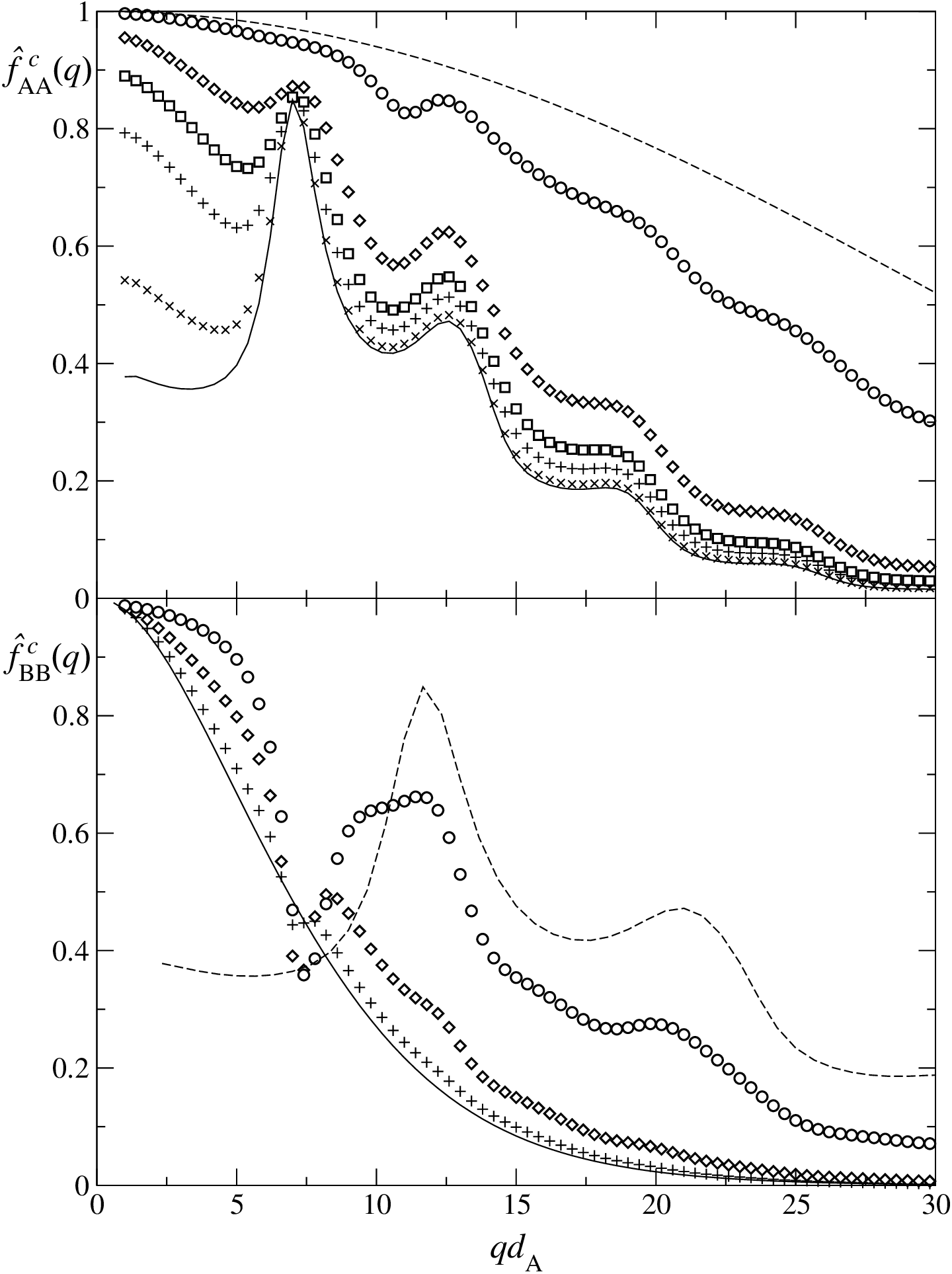}
\caption{\label{fig-fc06}
  Same as Fig.~\protect\ref{fig-fc08}, but for size ratio
  $\delta=0.6$.
}
\end{figure}

The above argument only depends on the fact that $\delta<1$, but not on the
precise ratio of localization lengths. Thus it is quite
general in binary HSM.
Fig.~\ref{fig-fc06} shows the scenario for $\delta=0.6$, i.e.\ for a larger
size disparity, and indeed one recognizes the same trends as above.
Here, the deviations of $\hat f^c_{\text{BB}}(q)$ from the tagged particle's
$f^{s,c}_{\text{B}}(q)$
set in faster with increasing $\hat x_{\text{B}}$ than it was
the case for $\delta=0.8$.
But one has to keep in mind that for smaller $\delta$, the changes in
$\hat x_{\text{B}}$ induce larger changes in the number concentration
$x_{\text{B}}$, cf.\ Eq.~(\ref{numberconc}).
The description of $\hat f^c_{\text{AA}}(q)$
as a simple interpolation between $f^{s,c}_{\text{A}}(q)$
and $f^c(q)$ as explained above
is notably worse, indicating that this simple picture quantitatively
only works for $\delta$ not too different from unity.
Also more pronounced in this case
are the changes in $\hat f^c_{\text{BB}}(q)$ for small $q$, going back to
the same reason as outlined above.
Let us note in addition that for both $\delta$, the
trend noticed for the diagonal elements is also found for
$\tilde f_{\text{AB}}(q)=F_{\text{AB}}(q)/S_{\text{AB}}(q)$, provided one is
sufficiently far away from those $q$ where a divergence
due to vanishing $S_{\text{AB}}(q)$ occurs.

\begin{figure}
\includegraphics[width=\figwidth]{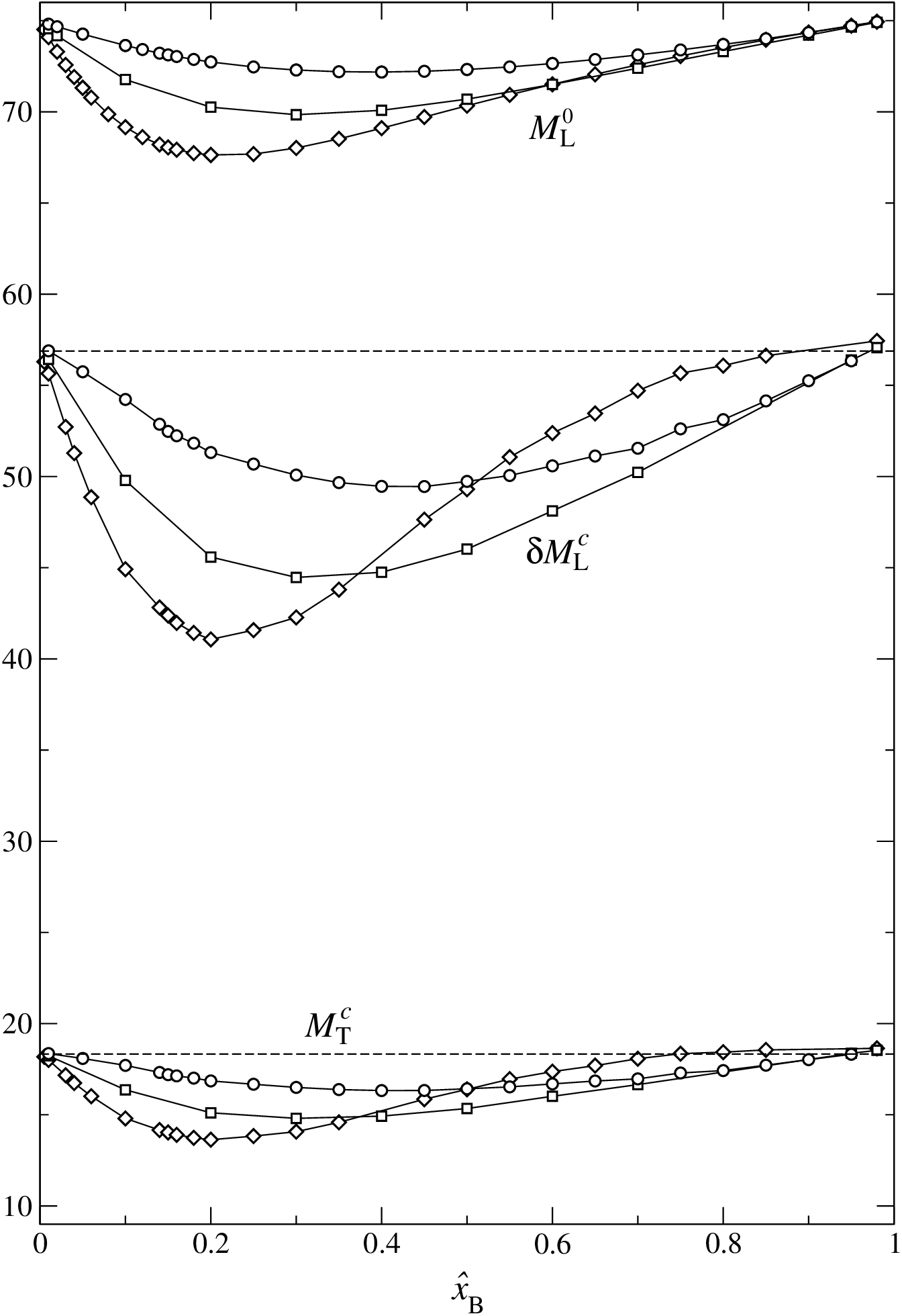}
\caption{\label{moduli}
  Isothermal longitudinal elastic modulus $M^0_{\text{L}}$, increase of the
  longitudinal elastic modulus at the transition points $\delta M_{\text{L}}^c$,
  and transversal
  elastic modulus $M_{\text{T}}^c$ at the liquid-glass transition points
  in units of $\varrho k_{\text{B}}T$
  as functions of the packing contribution $\hat x_{\text{B}}$ of the
  smaller particles for size ratios $\delta=0.8$ (circles), $0.7$
  (squares), and $0.6$ (diamonds).
  The dashed lines indicate the corresponding ideal mixing values
  evaluated for the one-component system.
}
\end{figure}

Macroscopic mechanic stability of the system is characterized
by the elastic moduli. The liquid exhibits a longitudinal elastic modulus given
by the structure factor through $M_{\text{L}}^0=\varrho(k_{\text{B}} T)
\sum_{\alpha\beta}x_\alpha S^{-1}_{\alpha\beta}(q\to0)x_\beta$
\cite{Hansen1986}. In the glass, the longitudinal modulus $M_{\text{L}}$
is larger, $M_{\text{L}}=M_{\text{L}}^0+\delta M_{\text{L}}$, due to the
arrest of the structure, Eq.~(\ref{modl}).
Figure~\ref{moduli} shows the
results for the binary HSM at the transition points for $\delta=0.6$, $0.7$,
and $0.8$, and also the critical shear modulus, Eq.~(\ref{modt}).
All quantities are shown in units of
$(\varrho k_{\text{B}}T)$ in order to more clearly reveal the effect of
composition change. Note that the total density $\varrho$ of the system
increases and superimposes a rise in the moduli one could call an
``ideal mixing'' contribution. This ideal mixing value is
given by the one-component values, $\delta M^c_{\text{L}}\approx56.9$ and
$M^c_{\text{T}}\approx18.3$, shown through dashed lines in Fig.~\ref{moduli}.
At intermediate $\hat x_{\text{B}}$, strong deviations from ideal mixing
occur. For all $\delta$ investigated here, the moduli decrease below their
one-component values, indicating that the system becomes softer upon
addition of smaller spheres. The effect increases with decreasing $\delta$ and
it is of the order of $40\%$ for
$\delta=0.6$.
It is partly connected with a corresponding increase in compressibility,
$\kappa=1/M_{\text{L}}^0$.
Indeed, one observes for given $\delta$ minima in all three quantities at
roughly the same $\hat x_{\text{B}}$. Let us nevertheless point out that
apart from this thermodynamic contribution to the softening of the glass,
mode-coupling effects still are necessary to explain the moduli for
$\delta=0.6$. This can be inferred from the crossing of the
$M_{\text{T}}^c$- and $\delta M_{\text{L}}^c$-versus-$\hat x_{\text{B}}$ curves
that is absent in $M_{\text{L}}^0$.

\section{Dynamics}\label{sec.dyn}

Close to an ideal glass transition, the essential aspects of the dynamics
are described by a universal scenario. This scenario has
been discussed comprehensively for one-component systems \cite{Franosch1997}.
The results of Ref.~\cite{Franosch2002} assure that these
universal results
are shared by the dynamics of the HSM. In particular, a two-step decay process
arises, with plateau values given by the critical glass form factors,
$\hat f^c_{\alpha\beta}(q)$, and power-law relaxations towards and from the
plateau, governed by anomalous exponents.
If the total packing fraction is increased towards the critical value
$\varphi^c$ with other parameters kept fixed,
a drastic increase in the relaxation time $\tau_\alpha$ of the slowest decay
process is obtained that is typical for glass-forming liquids.

In this section, we shall focus on the general, but non-universal features
of the glassy relaxation in the binary HSM. To demonstrate the
changes induced by different compositions, let us investigate a horizontal
intersection of the transition diagram of Fig.~\ref{phasediag} and consider a
change of the composition $\hat x_{\text{B}}$ for the total
packing fraction fixed at $\varphi=0.515$, i.e., at a value slightly below the
glass transition of the one-component hard-sphere system.
As above, the two different general scenarios, large and small size disparity,
shall be demonstrated using the values $\delta=0.6$
and $\delta=0.8$, respectively.

\subsection{General Features}

\begin{figure}
\includegraphics[width=\figwidth]{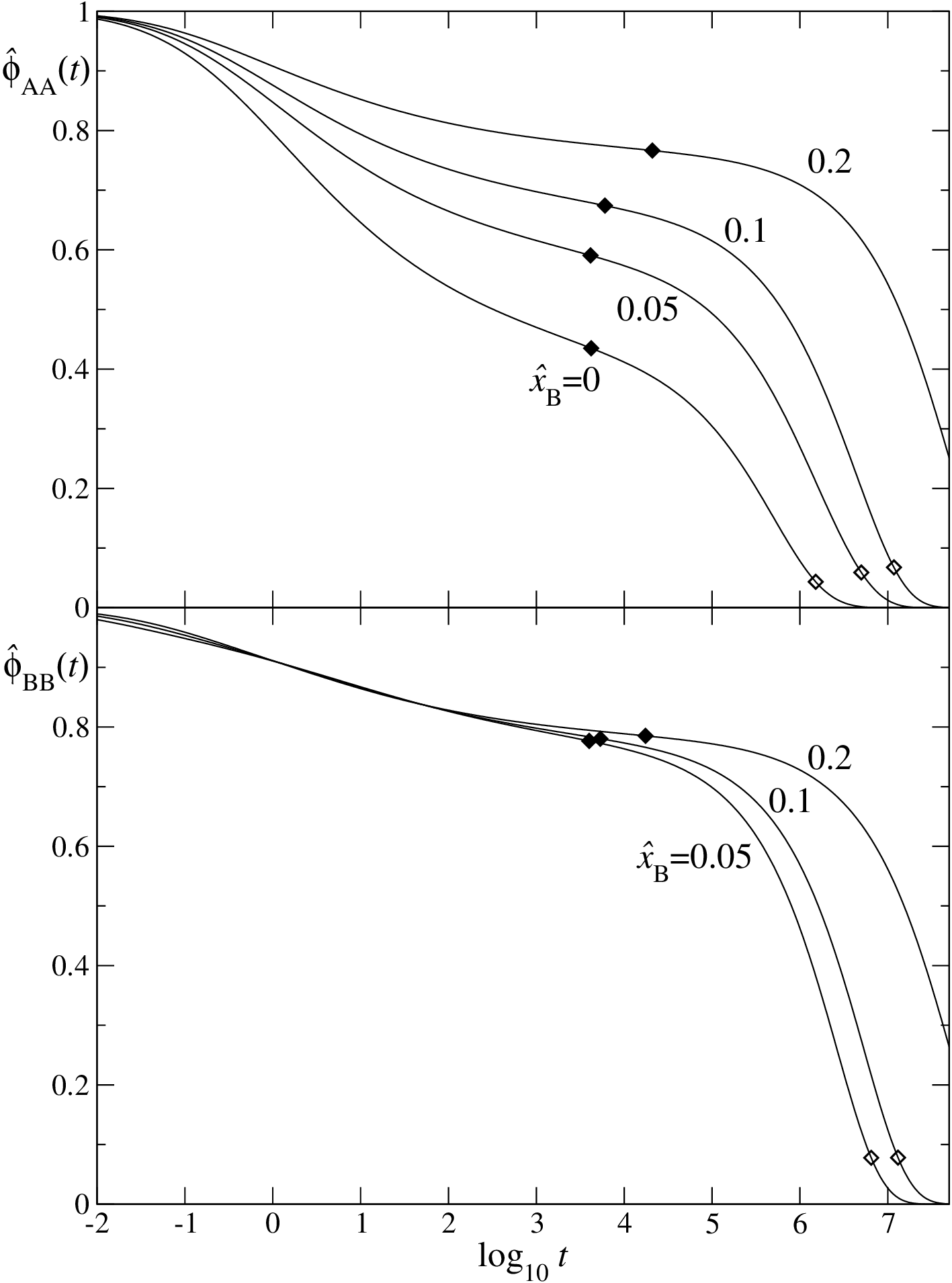}
\caption{\label{dyn08}
  Normalized partial density correlation functions $\hat\phi_{\alpha\alpha}(q,t)
  =\Phi_{\alpha\alpha}(q,t)/S_{\alpha\alpha}(q)$ for $\alpha=\text{A}$,
  $\text{B}$ of a binary HSM
  with size ratio $\delta=0.8$ and different packing contributions of the
  smaller particles $\hat x_{\text{B}}$
  for fixed packing fraction $\varphi=0.515$. The wave vector is
  $qd_{\text{A}}=5.4$.
  The unit of time here and in the following figures is chosen so that the
  short time diffusivity is $D^0_\alpha=0.01/d_\alpha$.
  Filled diamonds mark the intersection of the decay curves with the
  plateau value $\hat f_{\alpha\alpha}^c(q)$.
  The open diamonds mark $\alpha$-relaxation-time scales $\tau_{\alpha}(q)$
  defined by $\hat\phi_{\alpha\alpha}(q,\tau_\alpha(q))/
  \hat f^c_{\alpha\alpha}(q)=0.1$.
}
\end{figure}

\begin{figure}
\includegraphics[width=\figwidth]{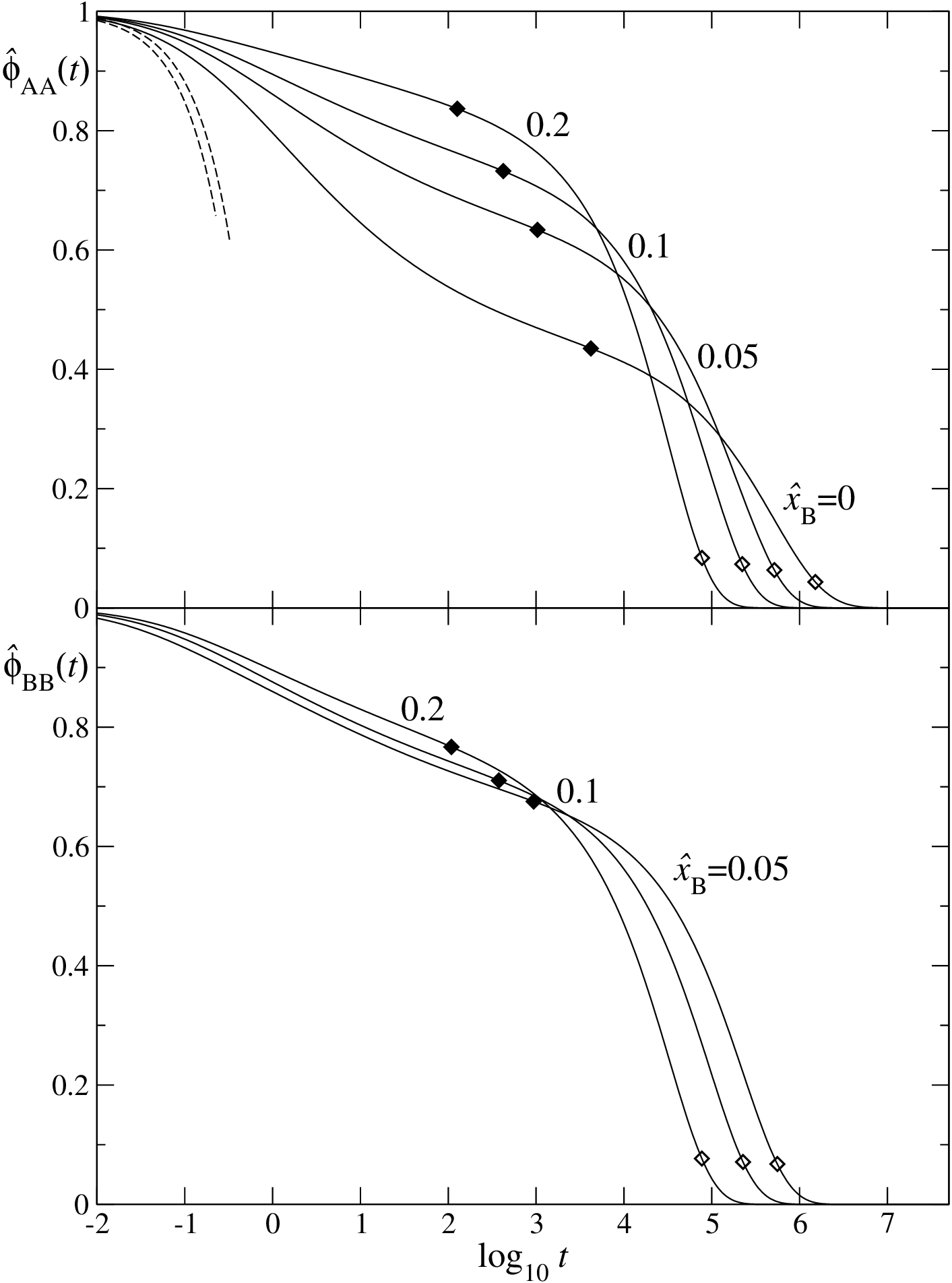}
\caption{\label{dyn06}
  Normalized partial density correlation functions
  $\hat\phi_{\alpha\alpha}(q,t)$
  as in Fig.~\protect\ref{dyn08}, but for the size ratio $\delta=0.6$.
  The dashed lines in the upper panel show the short-time approximation
  according to Eq.~(\protect\ref{diffusion-approx}) for $\hat x_{\text{B}}=0$
  and $0.2$, from left to right.
}
\end{figure}

The dynamics for $\hat x_{\text{B}}\le0.2$ is demonstrated by
Figs.~\ref{dyn08} and \ref{dyn06} for the $\text{AA}$ and $\text{BB}$
correlation functions.
We chose the wave vector $q=5.4/d_{\text{A}}$ below
the peak in $\hat f^c_{\text{AA}}(q)$; it corresponds roughly
to the one used in the light-scattering experiment of
Ref.~\cite{Williams2001b}.
The normalized correlators
$\hat\phi_{\alpha\alpha}(q,t)=\phi_{\alpha\alpha}(q,t)/S_{\alpha\alpha}(q)$
cross their plateau values $\hat f_{\alpha\alpha}^c(q)$ at certain
times, say $t_\alpha(q)$, $\hat\phi_{\alpha\alpha}(q,t_\alpha(q))=\hat
f_{\alpha\alpha}^c(q)$, that are marked by filled diamonds in the
figures. Close to the transition, the correlators are close to
this plateau for a large time interval. This is a manifestation of
the cage effect. In a leading order approximation for $\varphi^c-\varphi$
tending to zero, the time scale $t_\alpha(q)$ neither
depends on $\alpha$ nor on $q$ \cite{Goetze1991b}. The
independence of $\alpha$ is demonstrated to a good approximation
in the figures. As explained in connection with
Figs.~\ref{fig-fc08} and \ref{fig-fc06}, the plateau increases
with increasing $\hat x_{\text{B}}$ and the increase is more pronounced
for the larger majority particles $\text{A}$ than for the smaller
minority particles $\text{B}$.

The decay of the correlators below the plateau is referred to as
the $\alpha$ process. A characteristic time scale, $\tau_\alpha(q)$,
for this process shall be defined by specifying $90\%$ of the decay:
$\hat\phi_{\alpha\alpha}(q,\tau_\alpha(q))=0.1\,\hat f_{\alpha\alpha}^c(q)$.
These times are marked by open diamonds in the figures. For $\delta=0.8$,
Fig.~\ref{dyn08} demonstrates that the $\alpha$-relaxation scale increases with
increasing $\hat x_{\text{B}}$. This reflects the fact that with increasing
$\hat x_{\text{B}}$ the state corresponds to a smaller distance from the
transition point (compare Fig.~\ref{phasediag}).

The scenario for $\delta=0.6$, exhibited in Fig.~\ref{dyn06}, appears more
subtle. In this case, the glass-transition diagram implies the distance to the
transition to increase with increasing $\hat x_{\text{B}}$, thus leading to
faster decay on the $\alpha$-time scale. But the effect of increasing plateau
values with increasing $\hat x_{\text{B}}$ was seen above to occur for all
$\delta$.
The combination of these effects leads to a crossing of correlators, as
has also been observed in experiment \cite{Williams2001b}.

\begin{figure}
\includegraphics[width=\figwidth]{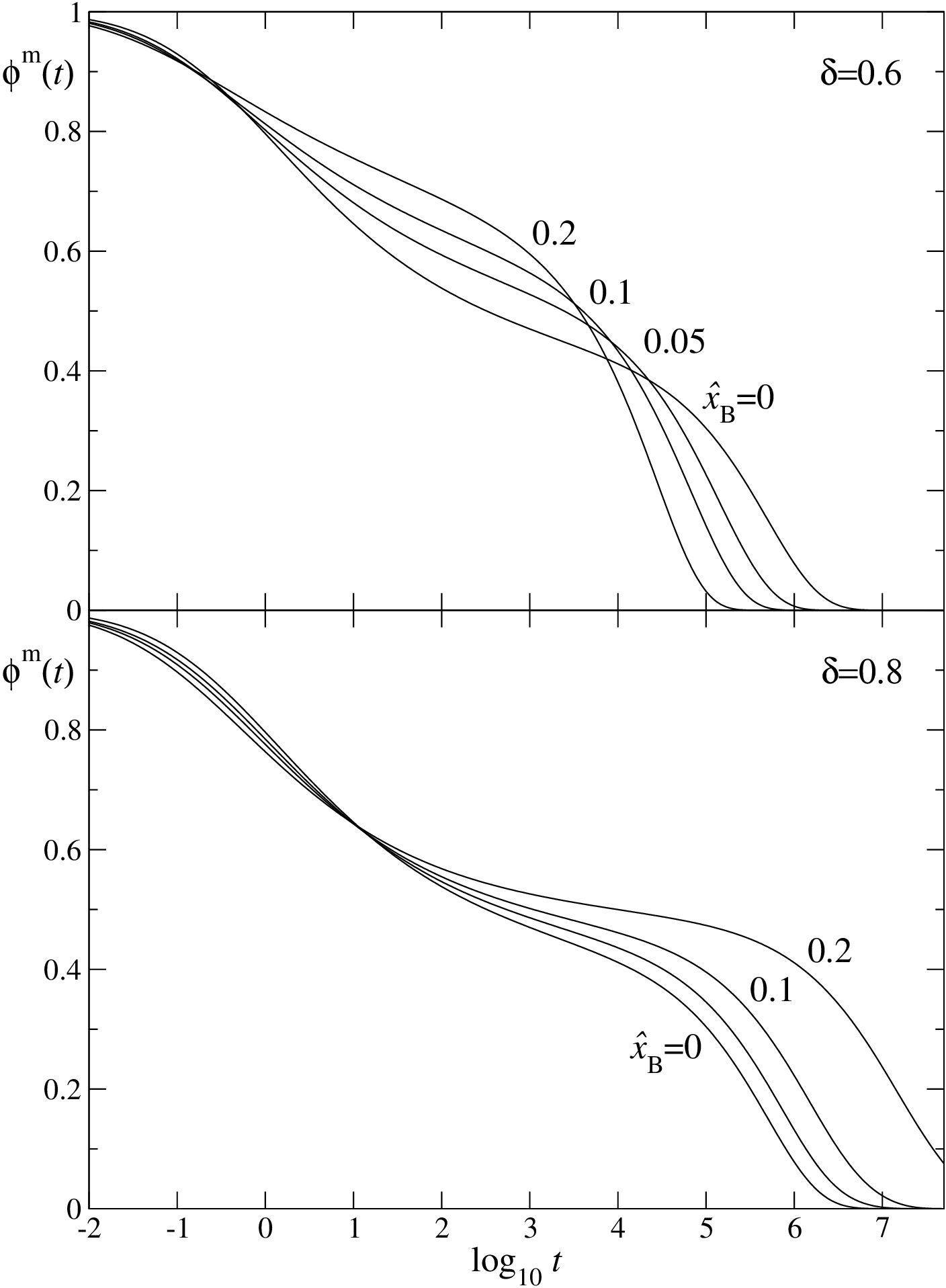}
\caption{\label{weighted-f}
  Sum $\phi^{\text{m}}(q,t)$ of the partial density correlation functions
  $\Phi_{\alpha\beta}(q,t)$ at wave vector $qd_{\text{A}}=5.4$, weighted
  according to Eq.~(\protect\ref{phiexp}) with scattering amplitudes
  $b_\alpha(q)$ as specified in Eq.~(\protect\ref{bij}).
  The packing fraction is kept constant at $\varphi=0.515$, and
  $\hat x_{\text{B}}=0$, $0.05$, $0.1$, $0.2$ as indicated by the labels.
  The upper panel shows the results for size ratio $\delta=0.6$, the lower
  one for $\delta=0.8$.
}
\end{figure}

In a dynamical light-scattering experiment one does not measure the
$\Phi_{\alpha\beta}(q,t)$ directly. Rather, one
measures a sum weighted with the scattering amplitudes $b_\alpha(q)$
\cite{Pusey1991},
\begin{equation}\label{phiexp}
  \phi^{\text{m}}(q,t)=\frac1{{\mathcal N}_q}\sum_{\alpha\beta}b_\alpha(q)
  b_\beta(q)\Phi_{\alpha\beta}(q,t)\,.
\end{equation}
Here, ${\mathcal N}_q$ is some normalization constant
chosen to satisfy $\phi^{\text{m}}(q,t=0)=1$.
It was a crucial point in Ref.~\cite{Williams2001b} to be able to vary the
$b_\alpha(q)$ without altering the dynamics. Thus, three
independent measurements of $\phi^{\text{m}}(q,t)$ could be used to invert
Eq.~(\ref{phiexp}) and therefore to determine the three distinct functions
$\Phi_{\alpha\beta}(q,t)$. The latter are better suited
for a comparison with the theory. But let us also demonstrate the dynamics for
a typical example of the directly measured function $\phi^{\text{m}}(q,t)$.
If one assumes the colloidal particles to be
uniform spheres, one gets \cite{Pusey1991}
\begin{equation}\label{bij}
  b_\alpha(q)\propto\frac{d_\alpha^3}{(qd_\alpha)^3}
    \left(\sin(qd_\alpha/2)-\frac{qd_\alpha}{2}\cos(qd_\alpha/2)\right)\,.
\end{equation}
Figure~\ref{weighted-f} shows the results for $\delta=0.6$
and $\delta=0.8$ at $qd_{\text{A}}=5.4$.
The same qualitative picture as discussed above for the
$\hat\phi_{\text{AA}}(q,t)$ correlator arises, yet the increase in plateau
values is less pronounced. The reason is a destructive interference effect
in Eq.~(\ref{phiexp}) caused by $\Phi_{\text{AB}}(q,t)\le0$.
This holds especially for $\delta=0.8$,
and also for smaller wave vectors.
Nonetheless, some increase remains in all cases, and one should be able to see
this in experiment. One could be tempted to analyze such data in terms of a
one-component model. However, this would be misleading. For a one component
system, the observed increase of the plateau
$f^{\text{m},c}(q)=\sum_{\alpha\beta}b_\alpha(q)b_\beta(q)F_{\alpha\beta}^c(q)$
would imply that the system becomes
stiffer upon increasing the contribution of smaller particles. But we have
seen above from a discussion of the mechanical moduli that the opposite is the
case.

\begin{figure}
\includegraphics[width=\figwidth]{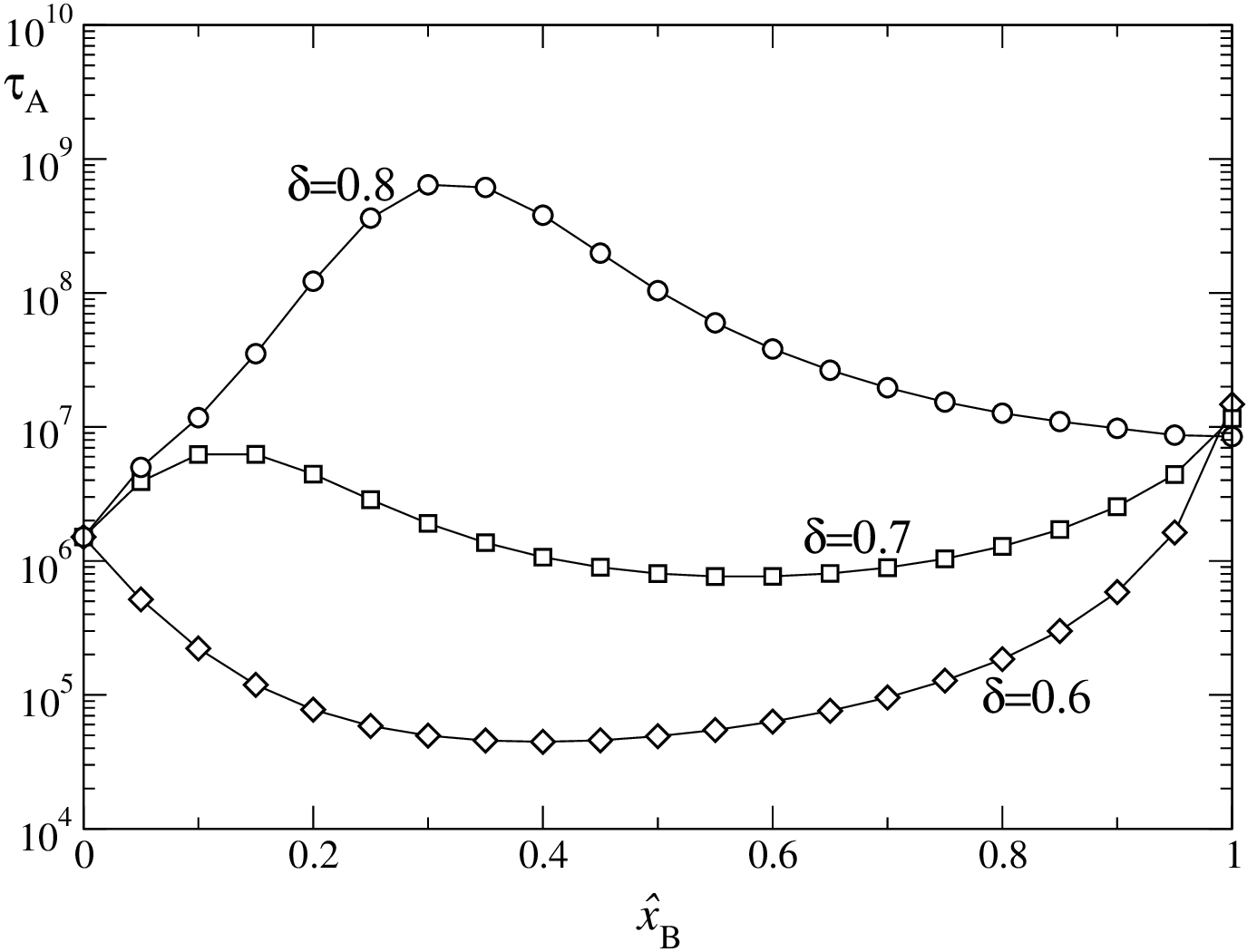}
\caption{\label{tau_alpha}
  $\alpha$-time scale $\tau_{\text{A}}$ defined through
  $\hat\phi_{\text{AA}}(q,\tau_{\text{A}})=0.1\,\hat f_{\text{AA}}^c(q)$
  for $qd_{\text{A}}=5.4$ at packing fraction
  $\varphi=0.515$, and $\delta=0.8$ (circles), $0.7$ (squares), and
  $0.6$ (diamonds). The lines are guides to the eye.
}
\end{figure}

Figure \ref{tau_alpha} exhibits $\alpha$-relaxation scales
$\tau_{\text{A}}$ for the larger particles as a function of mixing.
It corroborates the picture suggested by the
glass-transition diagram. Since the
$\alpha$ relaxation close to the glass transition
varies as $\tau_\alpha\sim(\varphi^c-\varphi)^{-\gamma}$, $\gamma>2.5$,
the variations of $\tau_{\text{A}}$
are much more pronounced than those of $\varphi^c$.
Note that the values of $\tau_{\text{A}}$ for different $\delta$ do not
necessarily coincide at $\hat x_{\text{B}}=1$.
MCT predicts all $\alpha$-relaxation times $\tau_\alpha(q)$ to be coupled.
Thus, the qualitative picture demonstrated in Fig.~\ref{tau_alpha} will also
hold for the $\alpha$-relaxation scales of other experimental quantities such
as the viscosities or inverse diffusivities.
Nucleation rates are also affected by the diffusivities; thus
Fig.~\ref{tau_alpha} demonstrates a possible reason for nucleation in binary
mixtures to vary strongly with changes of the composition.

\subsection{Dynamics Close to the Plateau}\label{sec.dynplat}

One notices in Figs.~\ref{dyn08} and \ref{dyn06} a trend
for the relaxation
onto the plateau value. This part of the curve, which deals with
the onset of structural relaxation, displays a slowing down of the
relaxation with increasing $\hat x_{\text{B}}$ for both cases considered for
$\delta$.
In principle, the relaxation in this time window is a result of both
structural and transient relaxation. In a leading approximation the
latter is given by
\begin{equation} \boldsymbol\Phi(q,t)=\exp\left[-q^2\boldsymbol D(q)t\right]
  \boldsymbol S(q)\,,
\end{equation}
with the matrix $\bs D(q)$ of short-time collective diffusion constants,
$\boldsymbol D(q)=(q^2\bs\tau(q))^{-1}\bs S(q)^{-1}$.
In particular, for a binary mixture this yields
\begin{equation}\label{diffusion-approx}
  \hat\phi_{\text{AA}}(q,t)=1-q^2 D'(q)t+{\mathcal O}(t^2)\,,
\end{equation}
where $D'(q)=x(q)D(q)$ with $x(q)=S(q)/S_{\text{AA}}(q)$, and $D(q)$ and $S(q)$
are the diffusion constant and the structure factor of the one-component system,
respectively.
It has already been noticed in Ref.~\cite{Williams2001b} that $x(q)<1$ for
small $q$. Thus, one expects a slowing down of the short-time diffusion due
to mixing in the limit of small $q$. For the wave vector discussed here,
however, the effect is small: with $\delta=0.6$, $\varphi=0.515$,
and $qd_{\text{A}}=5.4$, one gets $x\approx0.82$ ($0.76$, $0.78$) for
$\hat x_{\text{B}}=0.05$ ($0.1$, $0.2$).
The approximations resulting from
Eq.~(\ref{diffusion-approx}) are shown for $\hat x_{\text{B}}=0$ and $0.2$
as dashed lines in Fig.~\ref{dyn06}. One infers that this describes the
dynamics only for $\hat\phi_{\text{AA}}(q,t)\ge0.98$. Note that $x(q)$ is not
monotonous in $\hat x_{\text{B}}$, but the mentioned increase in the
stretching of the short-time relaxation with increasing $\hat x_{\text{B}}$ is.
Furthermore, at still larger wave vectors, one has $x(q)>1$ as
$x(q\to\infty)=1/x_{\text{A}}$, yielding faster short-time diffusion in the
mixture. But the slowing down of the relaxation towards the plateau persists
also for large $q$, as can be inferred from the numerical solutions.
Thus we conclude that the change in the short-time diffusion coefficients
is not sufficient to explain the observed effect.

\begin{figure}
\includegraphics[width=\figwidth]{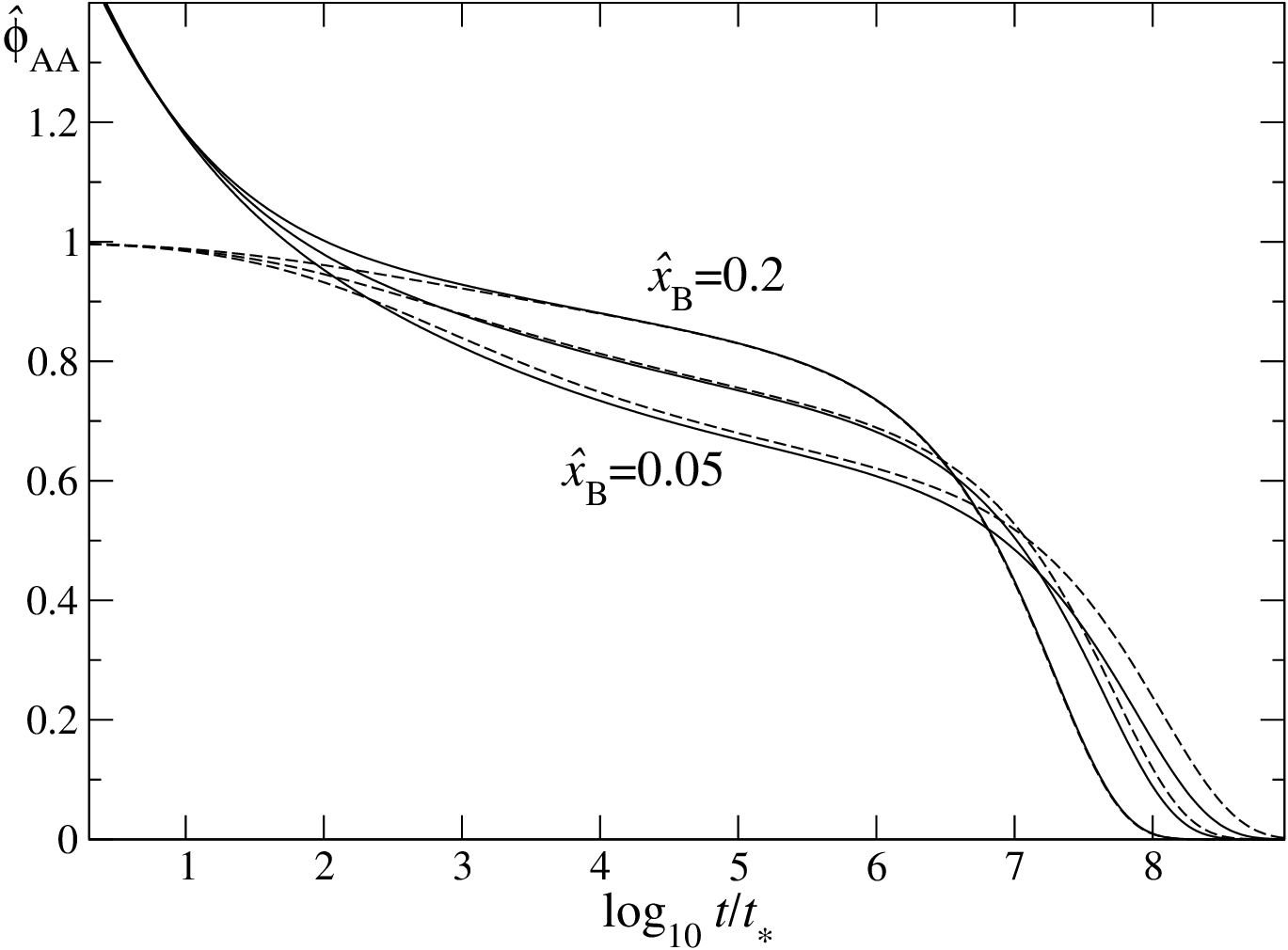}
\caption{\label{figstruc}
  Structural relaxation dynamics (solid lines) as defined by
  Eqs.~(\protect\ref{mctstruc}) for a binary HSM
  with $\delta=0.6$ and $\varphi=0.515$, with small particle
  packing contributions $\hat x_{\text{B}}=0.2$, $0.1$, and $0.05$ as indicated.
  The dashed lines are the solutions for the same parameters
  of the general MCT equations, Eqs.~(\ref{mcteq}), with the time
  scaled to match the structural-relaxation solution at long times for
  $\hat x_{\text{B}}=0.2$.
}
\end{figure}

Let us now focus on the structural relaxation as defined in
Sec.~\ref{sec.eom}A.
Figure~\ref{figstruc} presents solutions of Eqs.~(\ref{mctstruc}) for
$\delta=0.6$ and different
$\hat x_{\text{B}}$ at fixed $\varphi$, together with the solutions
reproduced from Fig.~\ref{dyn06}. The long-time parts of corresponding curves
can be scaled on top of each other, since there the dynamics depends on the
short-time behavior only through a scaling time $t_0$, as is demonstrated
for the $\hat x_{\text{B}}=0.2$ curve. For other
$\hat x_{\text{B}}$, the same observation is valid. Nevertheless, we have
applied the same rescaling as used for $\hat x_{\text{B}}=0.2$ instead of
matching $t_0$ and $t_*$ independently for different $\hat x_{\text{B}}$.
This is done in order to also demonstrate the drift of
the scaling time $t_0(\hat x_{\text{B}})$ with
$\hat x_{\text{B}}$. At short times, all structural relaxation curves follow the
same asymptote $t^{-1/3}$, and one notices that they deviate from one another
at roughly $t=10\,t_*$.
This demonstrates that the increase of the stretching in the initial decay with
increasing $\hat x_{\text{B}}$,
exhibited for $2.5\lesssim\log_{10}(t/t_*)\lesssim5.5$,
is a result of structural relaxation rather than transient dynamics.

In order to achieve a deeper understanding of the conclusions concerning the
initial part of the structural relaxation processes demonstrated above,
recall that one can derive scaling laws for an
analytical description of the correlators near their plateau values.
This has been discussed comprehensively for the one-component system
in Ref.~\cite{Franosch1997}. The theory is based on the observation that
the liquid-glass transition is described by a $A_2$ bifurcation of
Eq.~(\ref{mctfeq})
for the glass form factors. It is straightforward
to generalize the theory for one-component
systems to the case of interest here \cite{tvphd}. Let us
merely note the basic results
necessary to understand the following figures.

For the asymptotic expansion, one identifies a small parameter $\sigma$
and, connected to it, a time scale $t_\sigma=t_0|\sigma|^{-1/(2a)}$.
Here, the critical exponent $0<a<1/2$ is one of the nontrivial exponents of
MCT that is calculated
from the mode-coupling functional at the transition via the so-called
exponent parameter $\lambda$, $\lambda=\Gamma(1-a)^2/\Gamma(1-2a)$.
The separation parameter $\sigma$ is also calculated from the mode-coupling
functional, and is a smooth function of the control parameters that
vanishes at the transition. The conditions $\sigma>0$ and $\sigma<0$
characterize glass states and liquid states, respectively.
Setting $\hat t=t/t_\sigma$, one obtains an expansion in the small
quantity $\sqrt{|\sigma|}$.
\begin{multline}\label{asyeq}
  \bs\Phi(q,t)-\bs F^c(q)=\bs H(q)\sqrt{|\sigma|}g(\hat t)\\
  +\bs H(q)\left[|\sigma|h(\hat t)+\sigma\nu\right]
  +\bs K(q)\left[|\sigma|g(\hat t)^2-\sigma/(1-\lambda)\right]\\
  +\bar{\bs K}(q)\sigma/(1-\lambda)
  +{\mathcal O}(|\sigma|^{3/2})\,.
\end{multline}
Let us first explain the leading order contribution, which is
given by the first line
of Eq.~(\ref{asyeq}). It demonstrates the so-called factorization theorem,
in that it splits the wave-vector and control-parameter dependence off
from the time-dependence.
It is the critical eigenvector $\bs H(q)$ introduced above that governs
the former.
The latter is given by a master
function $g(\hat t)$ that is the solution of
\begin{equation} \frac{d}{d\hat t} \int_0^{\hat t}g(\hat t-t')g(t')\,dt'
  =\lambda g(\hat t)^2+\sgn\sigma\,,
\end{equation}
obeying $g(\hat t\to0)\sim(t/t_0)^{-a}$. The shape function
$g(\hat t)$ does not depend on the details of the mode-coupling
vertices, but only on the exponent parameter $\lambda$. Thus, the
factorization theorem predicts that all correlators for all models
resulting in the same $\lambda$ can
be rescaled to have the same shape, given by the master function $g(\hat t)$.

The corrections to the specified scaling law are given by the
terms of order $\sigma$ in Eq.~(\ref{asyeq}). They consist of a part parallel
to the critical
amplitude $\bs H(q)$, where a new correction-to-scaling shape function,
$h(\hat t)$, and a constant determined by the details of the mode-coupling
vertices, $\nu$, appear.
In addition, two correction amplitudes, $\bs K(q)$
and $\bar{\bs K}(q)$, to be evaluated from the mode-coupling functional,
are introduced by the next-to-leading order.
They explain that factorization holds with
different quality for different correlators.
One finds the $K_{\text{AA}}(q)$ and $\bar K_{\text{AA}}(q)$
to show the same qualitative variation with $q$
in the HSM considered here as in the one-component case discussed in
Ref.~\cite{Franosch1997}.
The only parameter that cannot be calculated within this approach is
the time scale $t_0$; it is fixed by matching
the long-time limit of the asymptotic solution at the critical point,
$\bs\Phi^c(q,t)=\bs F^c(q)+\bs H(q)(t/t_0)^{-a}+
{\mathcal O}(t^{-2a})$, to the
numerical solution at long times.

\begin{figure}
\includegraphics[width=\figwidth]{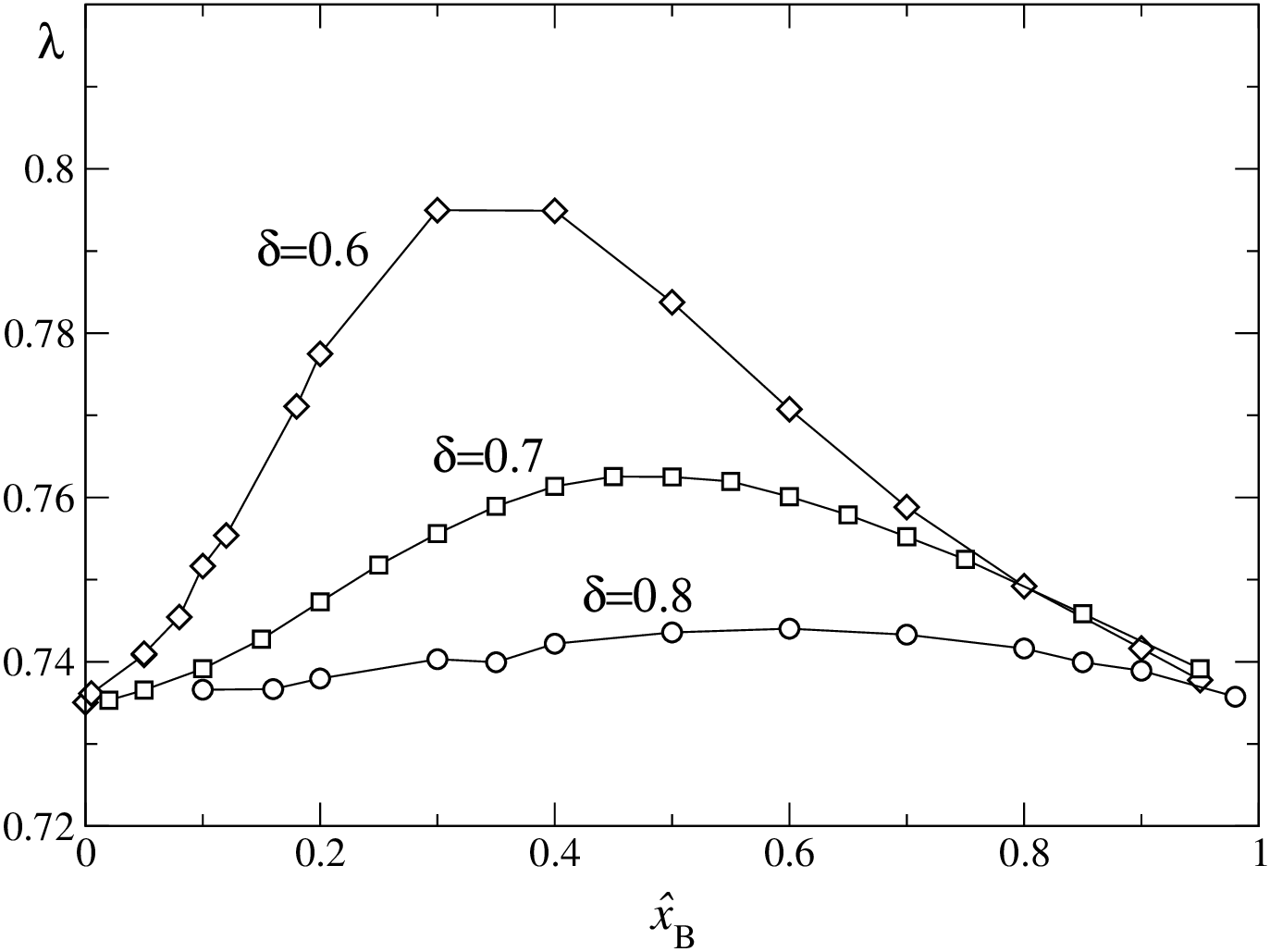}
\caption{\label{lplot}
  Exponent parameters $\lambda$ corresponding to the points shown in the
  transition diagram, Fig.~\ref{phasediag}; symbols indicate $\delta=0.6$
  (diamonds), $0.7$ (squares), and $0.8$ (circles). The lines are
  guides to the eye.
}
\end{figure}

We first investigate the variation of $\lambda$ as a function of the
composition, shown in Fig.~\ref{lplot}.
The exponent parameter is larger than the value found for the
pure hard-sphere system,
$\lambda(\hat x_{\text{B}}=0)=\lambda(\hat x_{\text{B}}=1)=0.736$. It
exhibits a maximum smaller than $0.8$ for $\delta\ge0.6$. As a result,
the critical exponent decreases relative to the value $a=0.311$ for the
hard-sphere system. In particular we get $\lambda=0.752$ ($0.778$),
and from this $a=0.304$ ($0.291$) for $\hat x_{\text{B}}=0.1$ ($0.2$).
As a consequence, the stretching of the decay towards the plateau increases
somewhat with increasing $\hat x_{\text{B}}$ and decreasing $\delta$. But this
effect is rather small and cannot explain the slowing down effect
specified above.

\begin{figure}
\includegraphics[width=\figwidth]{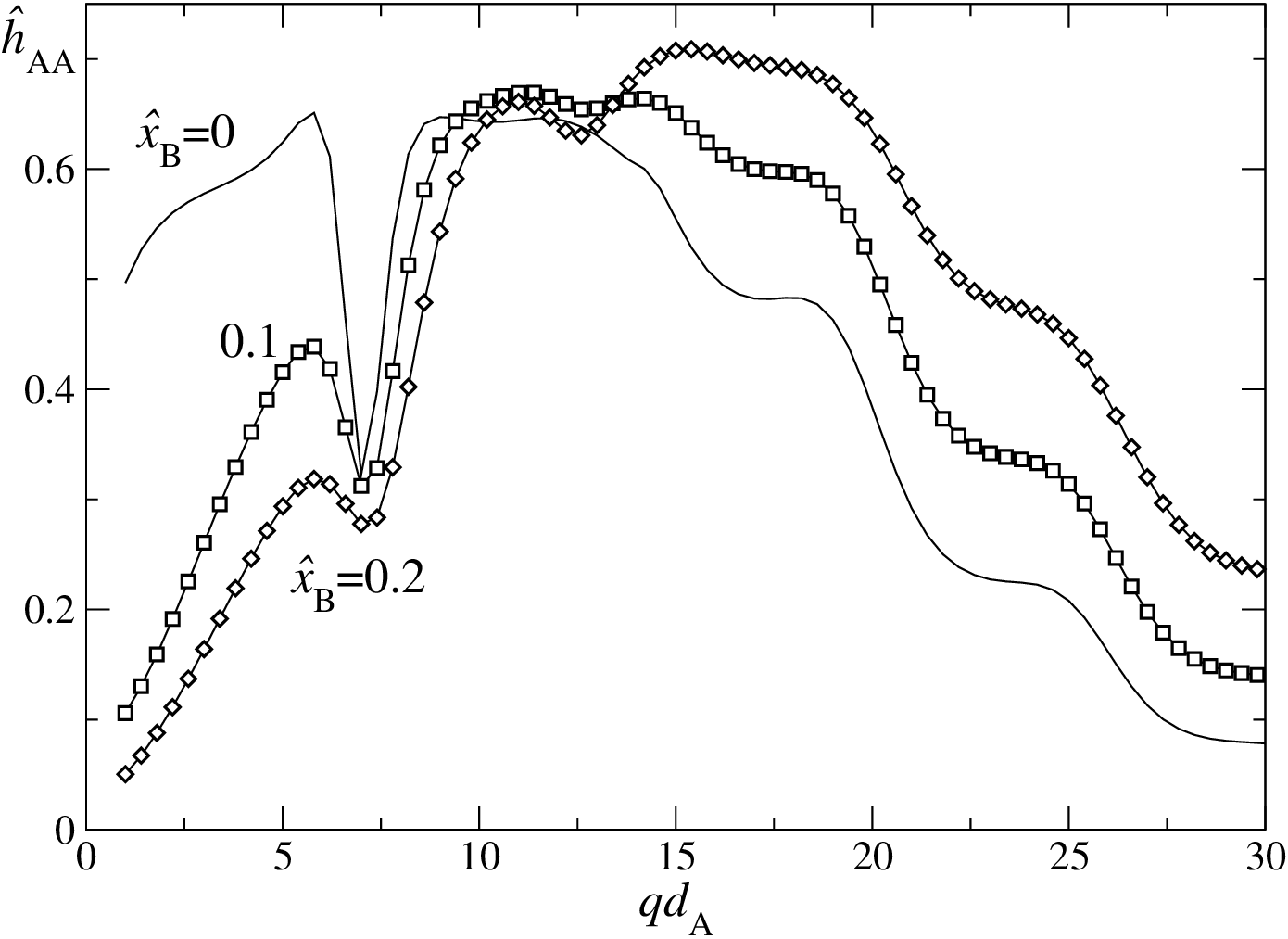}
\caption{\label{amplitudes}
  Normalized critical amplitudes $\hat h_{\text{AA}}(q)=H_{\text{AA}}(q)/
  S_{\text{AA}}(q)$ for $\delta=0.6$ and
  $\hat x_{\text{B}}=0.0$, $0.1$, and $0.2$ as indicated.
}
\end{figure}

Figure~\ref{amplitudes} shows the critical amplitudes $\bs H(q)$ in the case
$\delta=0.6$ for the
$\text{AA}$ correlator. The normalized quantity $\hat h_{\text{AA}}(q)
=H_{\text{AA}}(q)/S^c_{\text{AA}}(q)$ was chosen to match the representation
of Figs.~\ref{dyn08} and \ref{dyn06}.
While there is no general trend valid for
all $q$, we note that at the wave vector $q=5.4/d_{\text{A}}$ shown above,
$\hat h_{\text{AA}}(q)$
decreases significantly upon increasing $\hat x_{\text{B}}$. Let us emphasize
that the region of $qd_{\text{A}}\lesssim10$ is the one accessible in
dynamical light-scattering experiments on colloidal systems.
Furthermore, let us add that qualitatively the same change with
$\hat x_{\text{B}}$, although less pronounced, is observed in the $\delta=0.8$
case.
The decrease of $\hat h_{\text{AA}}$ yields a flattening of the
$\hat\phi(t)$-versus-$\log t$ curve within the time window that can be
described by the leading-order contribution to Eq.~(\ref{asyeq}).
The identified effect is further increased since the time
scale $t_0$ decreases with increasing $\hat x_{\text{B}}$. One gets
$t_0=0.4408$ ($0.2026$, $0.1385$) for
$\hat x_{\text{B}}=0$ ($0.1$, $0.2$) and other microscopic parameters as
given above.

\begin{figure}
\includegraphics[width=\figwidth]{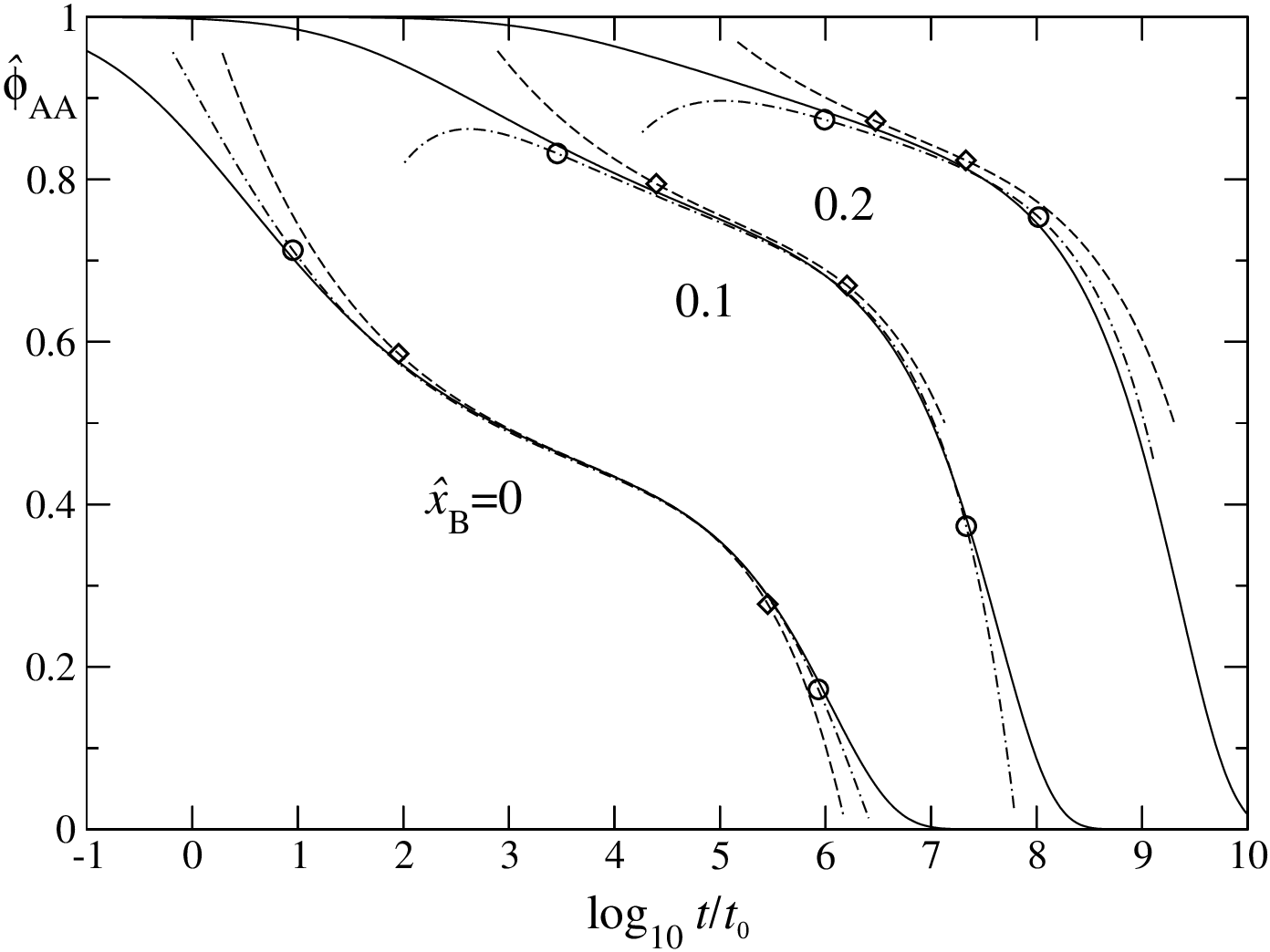}
\caption{\label{asyfit1}
  Asymptotic description of the normalized correlation functions
  $\hat\phi_{\text{AA}}(q,t)$ for $qd_{\text{A}}=5.4$,
  $\varphi=0.515$, $\delta=0.6$ and
  different $\hat x_{\text{B}}$ as indicated. The solid lines are the full
  solutions reproduced from Fig.~\protect\ref{dyn06}, but plotted as functions
  of $t/t_0$. The time scale $t_0$ is $0.4408$; $0.2026$ and $0.1385$ for
  $\hat x_{\text{B}}=0$, $0.1$, and $0.2$, respectively.
  Dashed and dash-dotted lines show the results of
  Eq.~(\protect\ref{asyeq}) up to order $|\sigma|^{1/2}$ and $|\sigma|$,
  respectively. The diamonds (circles) mark where the asymptotic solution
  up to leading (next-to-leading) order deviates by $0.01$ from the normalized
  correlator. Curves for $\hat x_{\text{B}}=0.1$ ($0.2$) have been
  translated along the $t$-axis by $2$ ($4$) decades for clarity.
}
\end{figure}

\begin{figure}
\includegraphics[width=\figwidth]{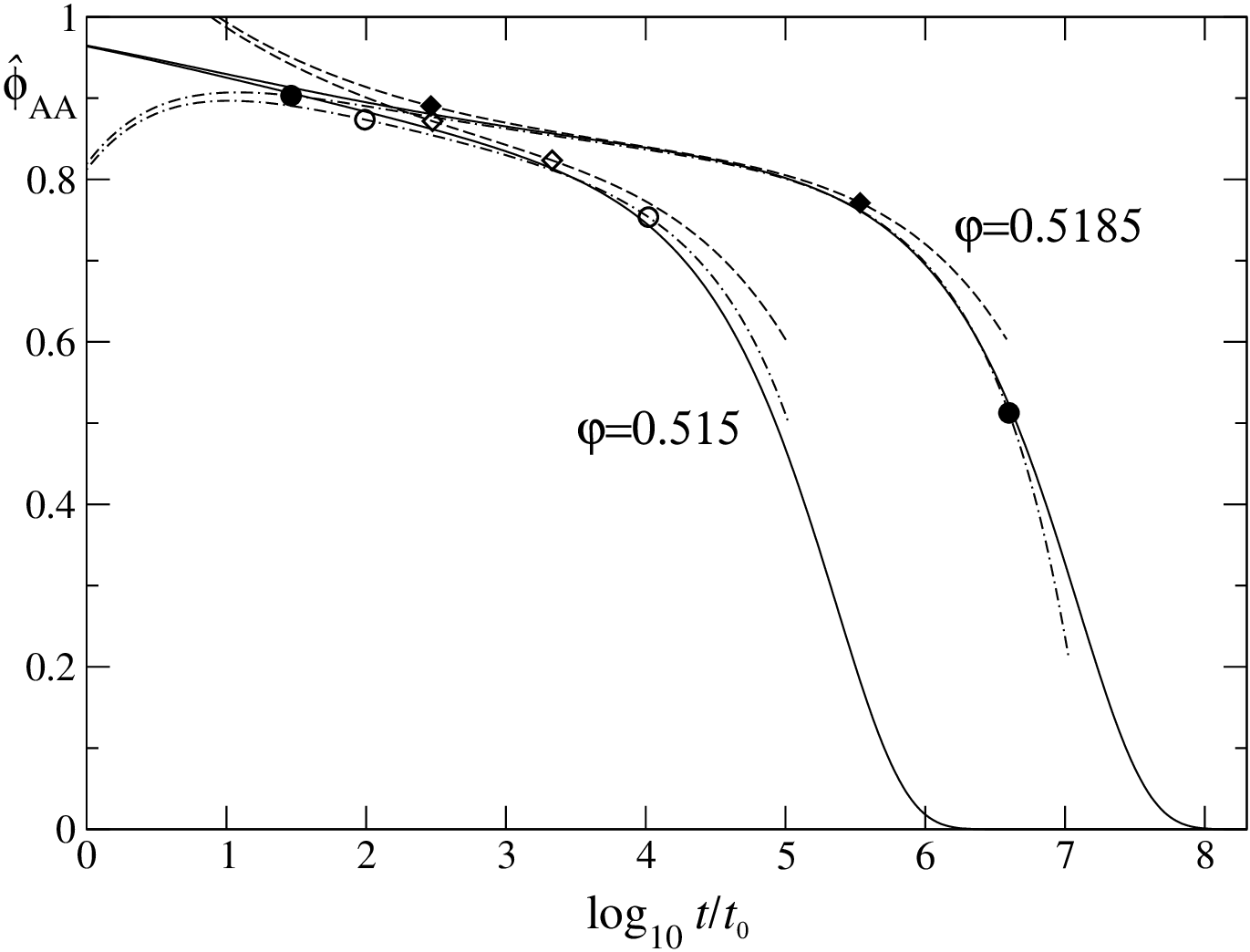}
\caption{\label{asyfit2}
  Asymptotic description of the normalized correlation functions
  $\hat\phi_{\text{AA}}(q,t)$ for $qd_{\text{A}}=5.4$,
  $\delta=0.6$ and $\hat x_{\text{B}}=0.2$
  at different $\varphi$ as indicated. Lines and symbols as in
  Fig.~\protect\ref{asyfit1}; open symbols refer to $\varphi=0.515$ and
  closed ones to $\varphi=0.5185$.
}
\end{figure}

Let us turn the preceding discussion into a quantitative demonstration by
comparing in Fig.~\ref{asyfit1} the asymptotic formula with the complete
solution for the $\hat\phi_{\text{AA}}$ correlator. The case
$\hat x_{\text{B}}=0$ shows a typical scenario for the one-component
system, where the first line of Eq.~(\ref{asyeq}) describes over $3$ decades in
time of the solution, as indicated by the open diamonds. This window of
the analytic description
is expanded by the next-to-leading-order formula by about $1$ decade
both at short and at long times, as can be seen from the circle symbols.
For $\hat x_{\text{B}}=0.1$ and $0.2$, the range of validity for both the
leading and the next-to-leading order is seen to shrink; at
$\hat x_{\text{B}}=0.2$ it is, including corrections, only about
$2$ decades. But to understand this, one has to keep in mind that the
distance from the critical point has increased by changing from
$\hat x_{\text{B}}=0$
to $\hat x_{\text{B}}=0.2$ for fixed total packing. Indeed,
we get $\sigma=-0.0027$ ($-0.0066$, $-0.011$) for $\hat x_{\text{B}}=0$
($0.1$, $0.2$), i.e.\ an increase in $\sigma$ by about a factor $4$.
Thus, the decreasing quality of the asymptotic description is merely due to an
increase of $\sigma$. This is corroborated
by Fig.~\ref{asyfit2}, where the $\hat x_{\text{B}}=0.2$ case is repeated
together with a point closer to the transition. For $\hat x_{\text{B}}=0.2$ and
$\varphi=0.5185$, the separation parameter is $\sigma=-0.0028$, similar to the
value discussed in Fig.~\ref{asyfit1} for $\hat x_{\text{B}}=0$.
Therefore, the regions of validity of the asymptotic expansions are
similar as well.
Indeed, the relevant quantity specifying the range of validity of the
asymptotic expansion is not the size of the logarithmic time interval,
but the size of the decay interval $|\hat\phi_{\text{AA}}(q,t)-
\hat f_{\text{AA}}^c(q)|$. Figures \ref{asyfit1} and \ref{asyfit2} demonstrate
that the analytic formula, Eq.~(\ref{asyeq}), describes the structural
relaxation in $\hat\phi_{\text{AA}}(q,t)$ towards the plateau
$\hat f_{\text{AA}}^c(q)$ below $0.70$, $0.85$, and $0.90$ for
$\hat x_{\text{B}}=0$, $0.1$, and $0.2$, respectively. This is the
regime where Fig.~\ref{dyn06} exemplifies the slowing down of this decay
with increasing $\hat x_{\text{B}}$ that was reported in
Ref.~\cite{Williams2001b}.

\section{Conclusions}\label{sec.conc}

Within the mode-coupling theory (MCT) for a binary hard-sphere
mixture, four mixing effects have been identified for states
near the ideal liquid-glass transition. First, mixing suppresses
intermediate-ranged ordering effects and this leads to an increase of the
small-wave-vector limit of the total structure factor,
Fig.~\ref{sqtotal}. Equivalently, the compression modulus of the
liquid decreases. A similar softening of the elastic restoring
forces is found for the moduli for compression and shear of the
glass near the transition points, Fig.~\ref{moduli}. Second, an
apparently opposite phenomenon is exhibited by the increase of the
Debye-Waller factors, i.e.\ a stiffening of the glass with respect
to spontaneous density fluctuations, Figs.~\ref{fig-fc08}
and \ref{fig-fc06}. This means primarily that mixing for fixed
packing leads to better
localization of the particles. The third effect is closely related
to this, viz.\ a stiffening of the cages of the localized particles
upon changes of composition. These changes are described
by the critical amplitude, which decreases upon mixing, as shown
in Fig.~\ref{amplitudes}. The universal MCT formula for the initial
part of the structural relaxation, Eq.~(\ref{asyeq}),
shows that this leads to a slowing down of the short-time
part of the glassy dynamics, as demonstrated in
Figs.~\ref{dyn08} and \ref{dyn06} and discussed quantitatively in
Figs.~\ref{asyfit1} and \ref{asyfit2}. The above described second and
third mixing effects have been identified originally in
experiments on colloids \cite{Williams2001b}.

The fourth general effect concerns the scale for the long-time
relaxation, i.e.\ the scale for hydrodynamic phenomena like
diffusion, or, more generally, for the $\alpha$-relaxation processes of
the liquid. Two scenarios are found as exhibited by the liquid-glass transition
diagram, Fig.~\ref{phasediag}, or by Fig.~\ref{tau_alpha}. For
small size disparity, mixing stabilizes the glass state. As
described above for the initial part of the structural relaxation,
also the final part of the decay is slowed down upon mixing. This
is shown in Fig.~\ref{dyn08} for the size ratio $\delta=0.8$.
However, for larger size disparities, an entropically induced
plasticizing effect is found. Due to mixing, the glass state is destabilized
and the $\alpha$-relaxation times decrease. As a result, the
$\hat\phi(t)$-versus-$\log(t)$ diagrams cross upon mixing as shown in
Fig.~\ref{dyn06} and observed for $\delta=0.6$ in the experiments
of Ref.~\cite{Williams2001b}.

In summary, our work demonstrates that MCT can explain
qualitatively the mixing effects on the glassy dynamics of
colloids observed for the size ratio $\delta=0.6$. A
quantitative comparison of the results of our theory with the data
of Ref.~\cite{Williams2001b} will be discussed in a subsequent
publication. Our theory suggests to also carry out experiments for
a size ratio near $\delta=0.8$ since a different scenario is
predicted for that case.

It can be expected that the results of our theory will also be of
some relevance to experiments on glass-forming binary metal alloys.
The formation of metallic glasses can to some extent be understood by treating
the constituent atoms as hard spheres, which will then be all of similar size
\cite{Meyer2002prea}. Even though in this paper we have dealt only with
Brownian short-time dynamics relevant for colloidal suspensions,
it is known that the long-time phenomena connected with the glass transition
are the same for Newtonian dynamics \cite{Franosch1998,Fuchs1999b}.
Recently, the concentration dependence of the critical temperature
$T_c$ was discussed for a
computer simulation of a $\mathrm{Co}_{100-x}\mathrm{Zr}_x$ model
\cite{Roessler2000}. This simulation used fine-tuned pair potentials to
model the metallic glass former; but if one estimates the size ratio of a
corresponding hard-sphere mixture from the atomic radii of $\mathrm{Co}$
and $\mathrm{Zr}$ \cite{Paszkowicz1988}, one gets $\delta\approx0.78$.
Indeed, in the simulation the curve $T_c(x)$ was found to have a maximum
at intermediate $x$, i.e.\ the glass transition was found to occur at smaller
coupling strengths. But this
corresponds to a decrease of $\varphi^c$ in the HSM
model. A similar reasoning holds for
computer-simulated $\mathrm{Ni}$-$\mathrm{Zr}$ melts \cite{Mutiara2001}.

Let us add some remarks on the results derived by Harbola and Das
\cite{Harbola2002}.
Their equations, as opposed to the ones studied in this paper, predict for the
glass-form factors $\hat f_{\alpha\beta}(q)$ and for the critical packing
fraction of the glass transition $\varphi^c$ a dependence on the mass ratio
$m_{\text{A}}/m_{\text{B}}$ of the two species. This result appears
surprising because one should not expect the equilibrium results for a
classical system to depend on the particles' inertia parameters. It is
obvious that the limit of vanishing concentration $x_{\text{B}}$ has to
reproduce the bell-shaped Lamb-M\"o{\ss}bauer factor for the glass-form
factor of the minority species, as discussed above in connection with
Fig. 5. This result is not obtained in the theory of Ref.~\cite{Harbola2002},
which, as a consequence, does not reproduce the experimental finding of an
increase in the correlator's plateau values upon mixing \cite{Williams2001b}.
Furthermore, the theory of Ref.~\cite{Harbola2002} predicts a much larger
increase of the critical packing fraction $\varphi^c$ upon mixing than measured
\cite{Williams2001b}. Indeed, it predicts that the glass transition can
disappear completely if the size ratio $\delta$ is smaller than a critical
value. This result seems implausible, since there is no obvious mechanism
which prevents the large particles from becoming a glass upon increasing the
density.

\begin{acknowledgments}
We thank W.~van~Megen, M.~Sperl, and S.R.~Williams for discussions and
valuable critique of our manuscript.
This work was supported by the Deutsche Forschungsgemeinschaft, through DFG
Grant Go.154/12-1.
\end{acknowledgments}

\bibliography{mct,add}
\bibliographystyle{apsrev}

\end{document}